\documentclass[pra,notitlepage,preprint,longbibliography]{revtex4-2}

\expandafter\let\csname equation*\endcsname\relax
\expandafter\let\csname endequation*\endcsname\relax
\expandafter\let\csname eqnarray*\endcsname\relax
\expandafter\let\csname endeqnarray*\endcsname\relax
\usepackage{array}
\usepackage{amsmath}
\usepackage{amssymb}
\usepackage{graphicx}
\usepackage{braket}
\usepackage{color}
\usepackage{float}
\usepackage{soul}
\usepackage{enumitem}
\usepackage[dvipsnames]{xcolor}
\usepackage{comment}
\usepackage{hyperref}
\hypersetup{colorlinks=true, citecolor=blue, urlcolor=blue, linkcolor=blue}
\usepackage{url}

\newcommand{\RN}[1]{\textup{\uppercase\expandafter{\romannumeral#1}}}

\renewcommand{\rm}{\mathrm}

\begin{document}
\setlength{\parindent}{0mm}

\title{Cavity optomechanics with ultra-cold Bose gases for quasiparticle state manipulation and prospects for sensing applications}

\author{Benjamin Maa{\ss}}
\author{Daniel Hartley}
\author{Kurt Busch}
\author{Dennis R\"atzel}
\email{dennis.raetzel@physik.hu-berlin.de}
\address{Humboldt-Universität zu Berlin, Institut für Physik, AG Theoretische Optik \& Photonik, Newtonstraße 15, 12489 Berlin, Germany}

\begin{abstract}
Ensembles of ultra-cold atoms have been proven to be versatile tools for high precision sensing applications. Here, we present a method for manipulation and readout of the state of trapped clouds of ultra-cold bosonic atoms. In particular, we discuss the creation of coherent and squeezed states of quasiparticles and the coupling of quasiparticle modes through an external cavity field. This enables operations like state swapping and beam splitting which can be applied to realize a Mach-Zehnder interferometer (MZI) in frequency space. We present two explicit example applications in sensing: the measurement of the healing length of the condensate with the MZI scheme, and the measurement of an oscillating force gradient with a pulsed optomechanical readout scheme. Furthermore, we calculate fundamental limitations based on parameters of state-of-the-art technology.

\end{abstract}

\maketitle

\section{Introduction}

The development of atom trapping and cooling technology has led to an explosion of applications in many areas of physics. One of the most well-known applications of control on the single atom scale is the atomic clock, which helped define the second in terms of the fundamental constants of nature \cite{ludlow2015optical}. Cold atoms are also used for quantum simulation, studying models from condensed matter \cite{lewenstein2007ultracold} as well as artificial gauge fields \cite{Dalibard2011coll} and other exotic topological states \cite{cooper2019topological}. Matter-wave interferometry with cold atoms has been applied to sensing applications such as the measurement of gravitational fields \cite{peters1999measurement,mcguirk2002sensitive,geiger2011detecting,bidel2018absolute} and precision tests of fundamental physics, such as measuring Newton's gravitational constant \cite{rosi2014precision}  and the fine structure constant \cite{parker2018measurement} and testing the equivalence principle \cite{Fray2004atomic,Schlippert2014quantum}, as well as physics beyond the standard model \cite{hamilton_atom-interferometry_2015}. 

When a three-dimensional cloud of atoms is cooled to such a degree that a macroscopic fraction of the atoms fall into the motional ground state, they condense into a Bose-Einstein condensate (BEC). 
Strictly speaking, in lower dimensions, Bose-Einstein condensation does not occur. Instead, quasi-condensates form that do not exhibit the long-rang order of a BEC. Unless we explicitly refer to BECs or quasi-condensates, we will use the term condensate for both in the following. Condensates of ultra-cold atoms can be used to push the sensitivity of the fundamental physics tests mentioned above even further \cite{Muentinga:2013int,van2010bose,gaaloul2014precision,Abend:2016atom,Hardman:2016sim,Asenbaum:2017pha}, even to applications in extraterrestrial space \cite{becker2018space,aveline2020observation}. The interaction between the atoms in the condensate leads to low-energy quasiparticles taking the form of phonons, i.e. quantised sound waves. Phonons are extensively studied in the field of quantum simulation \cite{bloch2012quantum,gring2012relaxation,rauer2018recurrences,Michael2019from}, analogue gravity including the simulation of event horizons \cite{barcelo2001analogue,lahav2010realization,steinhauer2014observation}, cosmic inflation \cite{fischer2004quantum} and gravitational waves \cite{bravo2015analog,hartley2018analogue}. Collective oscillations of condensates may also be used for sensing applications as demonstrated by the measurement of the thermal Casimir-Polder force presented in \cite{Obrecht:2007meas,Antezza:2004effect}. Further proposals include force sensing \cite{Motazedifard2019force}, gravimetry \cite{ratzel_dynamical_2018,bravo2020phononic}, tests of gravitationally induced collapse models \cite{Howl:2019expl} and even gravitational wave detection \cite{Sabin:2014gravwave,sabin_thermal_2016,Schuetzhold:2018int,Robbins_2019,robbins2021detection}.

Collective oscillations in BECs have been already studied in early experiments \cite{Jin:1996coll,Mewes:1996coll,Stamper-Kurn:1998coll} and it has been demonstrated that highly excited quasi-particle states can be created with light pulses \cite{Katz:2004high} and periodic modulations of the trap potential \cite{Jaskula:2012acoustic,Michael2019from}. Readout methods for quasiparticle excitations of condensates include self-interference of the Bose gas after release from the trap denoted as heterodyning \cite{Katz:2004high} or time-of-flight measurements (e.g. \cite{stamper1999excitation}) and in-situ phase contrast imaging \cite{Stamper-Kurn:1998coll,Schley:2013planck}. 

Here, we present an alternative approach
for creating, manipulating and reading out quasiparticle states of a condensate based on cavity optomechanics. The coupling of optical cavity modes to BECs has already been experimentally achieved \cite{brennecke2007cavity,brennecke2008cavity}. Our mechanism allows for displacement, single-mode squeezing or two-mode squeezing, and the coupling of two nearly arbitrary modes with interactions that are reminiscent of beam splitters and mirrors. Our manuscript is organized as follows: we introduce the cavity-condensate coupling in section \ref{sec:hamiltonian} and our proposed approach for state manipulation in section \ref{sec:stateman}. We present a Mach-Zehnder interferometer in frequency space in section \ref{sec:MZI} and a possible read-out mechanism via pulsed optomechanics in section \ref{sec:pulsed}. In section \ref{sec:damping}, we discuss damping mechanisms, and in section \ref{sec:applications}, we present two potential applications as examples of how to use our state manipulation scheme. We conclude our findings in section \ref{sec:conclusion}.

\section{Dynamics of the composite system}
\label{sec:hamiltonian}

\begin{figure}[b]
\centering
\includegraphics[width=6cm,angle=0]{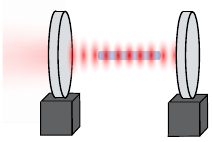}
\caption{\label{fig:cavity} {\small The setup considered in this paper is an ultra-cold atomic ensemble in an elongated trap which is effectively box-like in the elongated direction. The atomic ensemble is placed inside a Fabry--P\'{e}rot cavity to interact with a laser field in the cavity.     }}
\end{figure}

We begin by considering a condensate trapped in an external potential within a Fabry--P\'{e}rot type high finesse optical cavity (see  Figure \ref{fig:cavity}). The trap is considered to be elongated and 
the orientation of the cavity is considered to be aligned with the elongated axis of the condensate and the $z$-direction. Then, we restrict our considerations to quasiparticle modes in the elongated direction of the trap and treat the system in a one-dimensional way considering an effective cross sectional area $\mathcal{A}$ of the condensate \footnote{This approximation is valid if the parameters and the geometry of the setup are chosen such that the coupling of the modes in the elongated direction to those in the transverse directions is sufficiently suppressed, for example, in the case of very tight transverse confinement (as considered below in the example application).}.

The time evolution of the total system is described by the Hamiltonian $\hat{H}_\text{total}= \hat{H}_\rm{cav}+\hat{H}_\rm{disp}+\hat{H}_\rm{cond}$, where $\hat{H}_\rm{cav}$ is the cavity field Hamitonian, $\hat{H}_\rm{disp}$ is the coupling of light and atomic cloud that we will introduce below and the Hamiltonian that describes the time-evolution of the atomic cloud is
\begin{align}
    \nonumber \hat{H}_\rm{cond} = \int dz\,\hat{\psi}^\dagger\Bigg[ & -\frac{\hbar^2}{2m_a}\partial_z^2 + \frac{\tilde{g}}{2} \hat{\psi}^\dagger\hat{\psi}  + V_0 + \delta V_\rm{ext}\Bigg]\hat{\psi}
\end{align}
where $m_a$ is the atomic mass and $\tilde{g}$ is the atom-atom interaction strength. $V_0$ is the trap potential and $\delta V_\rm{ext}$ includes all other external potentials that may affect the condensate; for example, an external gravitational potential. We assume that $\delta V_\rm{ext}$ can be considered to be small in comparison to the trapping potential and only changes the density distribution of the condensate slightly. Later, we will consider the sensing of $\delta V_\rm{ext}$ via its effect on the condensate as a specific application.

The one-dimensional description represented by $\hat{H}_\rm{cond}$ can be directly derived from the three-dimensional standard description of interacting Bose gases \cite{Pitaevskii:2003bose}. For example, in the case of a three-dimensional condensate in a three-dimensional box trap with a box-shaped ground state \cite{gaunt2013bose}, $\tilde{g}=g/\mathcal{A}$ where $g=4\pi\hbar^2a_{sc}/m_a$ is the 3D coupling constant parameterised by the s-wave scattering length $a_{sc}$ \footnote{Note that $a_\rm{sc}$ can, in general, be widely tuned for some atom species that possess Feshbach resonances (e.g. $^{23}$Na \cite{inouye1998observation}, $^{85}$Rb \cite{roberts1998resonant} and $^{87}$Rb \cite{marte2002feshbach}) by employing strong magnetic fields to modify the s-wave scattering length. Unfortunately, also the three-body loss rate is strongly enhanced near a Feshbach resonance (see also \cite{chin2010feshbach}), where three-body loss is the dominating loss mechanism for trapped condensates and the dominating limitating factor for the maximal experimental time considered in this paper. Therefore, Feshbach resonances are of little use for our proposal and will not be considered.}.
Another possibility is to assume a strong harmonic transverse trap potential (such that the atoms are transversely mostly in the trap's ground state) which leads to a Gaussian shape of the condensate in the transverse direction and a one-dimensional quasi-condensate implying $\tilde{g}=g_\rm{1D}=g/(2\pi a_\perp^2)$, where $a_\perp=\sqrt{\hbar/(m_a\omega_\perp)}$ is the transverse oscillator length given by the transverse trapping frequency $\omega_\perp$. Then, the Hamiltonian (\ref{eq:hamiltonian}) is an approximation that corresponds to the case of a low one-dimensional density $\rho_\rm{1d} = \langle \hat{\psi}^\dagger\hat{\psi} \rangle$ of the condensate such that $\rho_\rm{1d} a_\rm{sc} \ll 1$ (see e.g. \cite{Salasnich2002eff,rauer2018recurrences}) \footnote{For $\rho_\rm{1d} a_\rm{sc} \gtrsim 1$, we would need to replace $\hat{\psi}^\dagger\hat{\psi}$ by more complicated functions of $\hat{\psi}^\dagger\hat{\psi}$ (see e.g. \cite{Salasnich2002eff,rauer2018recurrences}). For the sake of simplicity we refrain from this here.}.

\subsection{Light-matter coupling and full potential}

We consider a single optical cavity mode with annihilation and creation operators $\hat{a}$ and $\hat{a}^\dagger$ respectively, and free evolution with $\hat{H}_\rm{cav} = \hbar\omega_c \hat{a}^\dagger\hat{a}$ that is coupled to the atomic field operator $\hat{\psi}$ via the dispersive coupling Hamiltonian (photon absorption and stimulated re-emission) \cite{nagy2009nonlinear,ritsch2013cold}
\begin{align}\label{eq:Hdisp}
    \hat{H}_\rm{disp}= \int dz \,\hat{\psi}^\dagger(z)  \hbar\frac{g_0^2}{\Delta_A}f_\rm{cav}^2(z)\,\hat{a}^\dagger\hat{a}\,\hat{\psi}(z) \,.
\end{align}
To achieve this form of coupling, the cavity mode frequency is chosen close to an atomic resonance with a detuning $\Delta_A$ and the single photon Rabi frequency $g_0 = d\sqrt{\omega_c/(2 \hbar\epsilon_0\mathcal{V}_c)}$, where $d$ is the atomic dipole moment along the cavity mode polarization, $\mathcal{V}_c = \mathcal{A}_c \int dz \,|f_\rm{cav}(z)|^2$ is the effective cavity mode volume, $\mathcal{A}_c$ is the effective cross sectional area of the cavity mode and $f_\rm{cav}(z)$ is the cavity mode function.

We assume that the cavity mode is driven by a strong laser field with frequency $\omega=\omega_c +\Delta_c$, and we move into the corresponding rotating frame. This allows us to treat the cavity field perturbatively by splitting the mode operators $\hat{a}$ and $\hat{a}^\dagger$ into their expectation values and fluctuations as $\hat{a}=\langle \hat{a} \rangle + \delta \hat{a}$ \cite{aspelmeyer2014cavity,pikovski:thesis}. Since $\hat{H}_\rm{disp}$ is invariant under phase factors $\hat{a} \rightarrow \hat{a}e^{i\zeta}$, without loss of generality, we consider $\langle \hat{a} \rangle$ as real valued and set $\langle \hat{a} \rangle = \sqrt{N_{ph}}$, where $N_{ph}$ is the number of photons in the cavity mode. The photon number is related to the circulating power in the cavity as $P_c = \hbar \omega_c N_{ph} c/(2L_c) $, where $c$ is the speed of light and $L_c$ is the length of the cavity. $P_c$ can be varied by modulating the cavity pump power on time scales much larger than the life time of photons in the cavity mode. This is the basis for the state manipulation of the quasiparticles that we present in this work, and in the following, we consider $N_{ph}$ as generally time-dependent.

Introducing the split $\hat{a}=\langle \hat{a} \rangle + \delta \hat{a}$ into $\hat{H}_\rm{disp}$ and neglecting the second order term in $\delta \hat{a}$, we obtain \cite{aspelmeyer2014cavity}
\begin{align}\label{eq:hamiltonian}
    \nonumber \hat{H}_\text{total} = -\hbar\Delta_c \delta\hat{a}^\dagger\delta\hat{a} + \int dz\,\hat{\psi}^\dagger\Bigg[ & -\frac{\hbar^2}{2m_a}\partial_z^2 + \frac{\tilde{g}}{2} \hat{\psi}^\dagger\hat{\psi}  + V\left(z,t\right) \\
    &  +  \hbar\frac{g_0^2\sqrt{N_{ph}(t)} }{\Delta_A}f_\rm{cav}^2(z)\left(\delta\hat{a}^\dagger  + \delta\hat{a}  \right)  \Bigg]\hat{\psi}\,,
\end{align}
where the full potential acting on the condensate is
\begin{equation}\label{eq:potential}
	V\left(z,t\right) = V_0\left(z\right) + \delta V_\rm{ext}\left(z,t\right)   +  \hbar\frac{g_0^2}{\Delta_A}f_\rm{cav}^2\left(z\right)N_{ph}(t) = V_0\left(z\right) + \delta V\left(z,t\right) \,.
\end{equation}
We restrict our considerations to situations where $\delta V$ oscillates around a mean. We split $\delta V$ into its time-average $\overline{\delta V}$ and the purely oscillating part $V_\rm{osc}=\delta V - \overline{\delta V}$. Later we will assume that $V_\rm{osc}$ oscillates on resonance with a quasiparticle excitation or quasiparticle transitions. Therefore, we can already conclude that $\overline{\delta V}$ modifies the stationary condensate ground state and the quasiparticle mode functions, while $V_\rm{osc}$ drives the quasiparticle modes.

\subsection{Condensate ground state and Bogoliubov approximation}

We split the field operator into a part describing the atoms in the collective ground state of the atomic ensemble and a part describing trapped atoms in states above the ground state. In the Heisenberg picture, we split the atom field operator into a part describing the macroscopic condensate fraction and a field operator $\hat\vartheta$ that describes the non-condensate atoms in the form
\begin{eqnarray}\label{eq:expansion}
	\hat\psi(z,t)= (\hat c_0 \psi_0(z) + \hat\vartheta(z,t) )	e^{-i\left( \mu t + \int_0^t dt'\, \delta\mu_\rm{osc}(t')\right)/\hbar} \,,
\end{eqnarray} 
where $\hat c_0$ is the annihilation operator for atoms in the collective ground state $\psi_0$ which is normalized as $\int dz\, |\psi_0|^2 = 1$ and fulfills the stationary Gross-Pitaevski (GP) equation 
\begin{equation}\label{eq:statgp}
	\left(-\frac{\hbar^2}{2m} \partial_z^2  + V_0 + \overline{\delta V}  + \tilde{g}N_0 |\psi_0|^2 \right)\psi_0 = \mu\, \psi_0\,,
\end{equation}
with the chemical potential $\mu$. Furthermore, we have defined $\delta\mu_\rm{osc} = \int dz \, |\psi_0|^2\, V_\rm{osc}$ which can be interpreted as a time-dependent shift of the chemical potential due to the oscillating part of the external potential  (see Appendix \ref{sec:intham}).

Assuming that $\overline{\delta V}$ perturbs the ground state of the condensate only slightly, we can make the ansatz $\psi_0 = \bar\psi_0 + \delta \psi_0$ with $\bar\psi_0 $ a solution of Eq. (\ref{eq:statgp}) with $\overline{\delta V}\rightarrow 0$. Expressions for the perturbation $\delta \psi_0$ can be found as a solution of a linearized version of the GP equation as presented in Appendix \ref{sec:gstate_pert}. The perturbation leads to a modification of the quasiparticle modes which, in turn, leads to terms of second order in $\delta V$ in the quasiparticle Hamiltonian (see also Appendix \ref{sec:gstate_pert}) and we will neglect $\overline{\delta V}$ in the following (restricting our considerations to effects of first order in $\delta V$).

Since the time evolution of the field operator $\hat\psi(z,t)$ is governed by the Heisenberg equation with respect to the Hamiltonian $\hat{H}_\text{total}$, we find that the time evolution of $\hat\psi'(z,t) :=\hat c_0 \psi_0(z) + \hat\vartheta(z,t)$ is governed by the Heisenberg equation with respect to the Hamiltonian
\begin{equation}
	\hat H_\text{total}':= \hat H_\text{total} - (\mu+\delta\mu_\rm{osc}) \hat N
\end{equation} 
(similar to the grand canonical Hamiltonian) where $\hat N(t) = \int dz\, \hat\psi^{\prime\dagger}(z,t) \hat\psi^\prime(z,t)$ is the number operator of the atom field. 

In the next step, we apply the Bogoliubov approximation to $\hat H_\text{total}'$, where the field $\hat\vartheta$ is treated as a small perturbation while the ground state is strongly occupied such that the replacement $\hat c_0 \rightarrow \sqrt{N_0} \mathbb{I}$ can be performed, where $N_0$ is the number of atoms in the condensate. Then, we obtain the driving Hamiltonian (see Appendix \ref{sec:intham} for the details of the derivation)
\begin{eqnarray}\label{eq:Hprimeint}
	  \hat H_\rm{dr} & = &  \sqrt{N_0} \int dz\, (V_\rm{osc} - \delta\mu_\rm{osc})\left(\psi_0^* \hat\vartheta + \psi_0 \hat\vartheta^{\dagger}\right) + \int dz\, \hat\vartheta^{\dagger}(V_\rm{osc} -\delta\mu_\rm{osc})\hat\vartheta\,.
\end{eqnarray}
In lowest order in the atom field operators, the back action on the cavity mode fluctuation quadrature is given via the Hamiltonian
\begin{equation}\label{eq:backaction_hamiltonian_unexp}
	\hat H_\rm{ba} = \hbar\sqrt{N_0} \frac{g_0^2\sqrt{N_{ph}} }{\Delta_A} \left(\delta\hat{a}^\dagger  + \delta\hat{a}  \right)  \int dz\, f_\rm{cav}^2  \left(\psi_0^*\hat\vartheta + \psi_0\hat\vartheta^{\dagger}   \right)\,.
\end{equation}
Here, a term that is proportional to $\left(\delta\hat{a}^\dagger + \delta\hat{a}\right)$ and independent of $\hat\vartheta$ has been absorbed into a re-normalization of the cavity mode frequency \footnote{This is the effect of the refractive index change due to the presence of the condensate atoms in the cavity.} (see also Appendix \ref{sec:freqshift} for how to include this frequency shift from the start).

\subsection{Quasiparticle mode expansion}

We expand the field operator in terms of Bogoliubov modes describing the quasiparticle excitations as
\begin{equation}
	\hat\vartheta = \sum_n  \left(u_n\hat b_{n} + v_n^*\hat b_{n}^\dagger \right)\,,
\end{equation}
where $[b_{n},b_{m}^\dagger]=\delta_{nm}$ and the normalized mode functions $u_n$ and $v_n$ fulfill the stationary Bogoliubov-de~Gennes (BDG) equations which can be found in Appendix \ref{sec:intham}. The expansion of $\hat\vartheta$ in the Bogoliubov basis diagonalizes the free quasiparticle Hamiltonian $\hat H_\rm{BdG}$ to second order in $\hat\vartheta$ (the Bogoliubov-de~Gennes Hamiltonian, see Appendix \ref{sec:intham}), which governs the free evolution of the condensate with $V(z,t)\rightarrow V_0(z)$, that is, $\hat H_\rm{BdG} = \sum_n \hbar \omega_n \hat b_{n}^\dagger \hat b_{n}$.

Then, taking the free time evolution of the creation and annihilation operators $\delta\hat{a}^\dagger$, $\delta\hat{a}$, $\hat{b}^\dagger$ and $\hat{b}$ into account, we obtain the back-action Hamiltonian in the corresponding rotating frames (the interaction picture) as
\begin{align}
        \hat{H}_\rm{ba} & = \left(\delta\hat{a}^\dagger e^{-i\Delta_c t} + \delta\hat{a}  e^{+i\Delta_c t}\right)\sum_{n}\kappa_{n} \left(\hat{b}_{n}e^{-i(\omega_{n}t-\theta_n)}  +  \hat{b}_{n}^{\dagger}e^{i(\omega_{n}t-\theta_n)}\right)
    \label{eq:backaction_hamiltonian}
\end{align}
where the real quantities $\kappa_n$ and $\theta_n$ are defined such that
\begin{align}\label{eq:coupling}
	\kappa_n e^{i\theta_n} = \frac{\hbar g_0^2\sqrt{N_0 N_{ph}} }{\Delta_A }  \int dz \,f_\rm{cav}^2  \left( \psi_0^* u_n + \psi_0 v_n \right) \, \,.
\end{align}
Using the mode decomposition above, we find that the driving Hamiltonian in Eq. (\ref{eq:Hprimeint}) assumes the form
\begin{align}    \label{eq:interaction_hamiltonian}
        \nonumber \hat{H}_\rm{dr} & =\sum_{n} \left(P_{n} \hat{b}_{n} e^{-i\omega_{n}t}  + P_{n}^* \hat{b}_{n}^{\dagger}e^{i\omega_{n}t}\right)+\sum_{n}\left(O_{n}\hat{b}_{n}^{\dagger}\hat{b}_{n} + \left( N_{n}  \left(\hat{b}_{n}^{\dagger}\right)^{2}e^{+2i\omega_{n}t} + N_{n}^*\,\hat{b}_{n}^{2}e^{-2i\omega_{n}t}\right)\right)\\
        & \hphantom{==}+\sum_{n,l<n} \left( M_{nl} \hat{b}_{n}^{\dagger}\hat{b}_{l}e^{i(\omega_{n} - \omega_{l})t}  + M_{nl}^* \hat{b}_{l}^{\dagger}\hat{b}_{n}e^{-i(\omega_{n} - \omega_{l})t}  \right)  \\
        & \nonumber  \hphantom{==}+\sum_{n,l<n} \left(L_{nl} \hat{b}_{n}\hat{b}_{l} e^{-i(\omega_{n} + \omega_{l})t}  + L_{nl}^* \hat{b}_{n}^{\dagger}\hat{b}_{l}^{\dagger}e^{i(\omega_{n} + \omega_{l})t}  \right) ,
\end{align}
where the form of each of the time-dependent coefficients is
\begin{align}\label{eq:coefficients}
      \nonumber  P_{n} & = \sqrt{N_0}   \int dz \,  (V_\rm{osc} - \delta\mu_\rm{osc}) \,\left( \psi_0^* u_n + \psi_0 v_n \right) \\
       \nonumber O_{n} & = \int dz \, (V_\rm{osc} - \delta\mu_\rm{osc})\,  \left(|u_n|^2 + |v_n|^2\right)   \\
       N_{n} & =   \int dz \, (V_\rm{osc} - \delta\mu_\rm{osc})\,  u_n^* v_n^*   \\
        \nonumber  M_{nl} & =   \int dz \, (V_\rm{osc} - \delta\mu_\rm{osc})\,  \left( u_n^* u_l + v_n^* v_l\right)   \\
     \nonumber   L_{nl} & =  \int dz \, (V_\rm{osc} - \delta\mu_\rm{osc})\,  \left( u_n v_l + v_n u_l\right)     \,.
\end{align}
and we have neglected a vacuum term $\propto |v_n|^2$. The driving Hamiltonian is one of the main results of this work. The coefficients can be interpreted as the moments of the external potential with respect to the quasiparticle modes and different processes.  The first term with the coefficient $P_n$ corresponds to a linear displacement of the quasiparticle state by the cavity field or potential perturbation. The second order terms can be interpreted as various nonlinear processes familiar from optical four-wave mixing; the term with the coefficient $O_n$ corresponds to a time-dependent frequency shift of the quasiparticle modes, the terms with the coefficients $N_n$ and $L_{nl}$ take the form of one-mode and two-mode squeezing operators, and the term with the coefficient $M_{nl}$ has the form of a beam splitting operation.

\section{State manipulation by modulated cavity power}
\label{sec:stateman}

In the following, we consider a cavity field which is periodically intensity modulated on resonance with a particular quasiparticle mode or mixture thereof at frequency $\omega_m$. Then, particular processes in $\hat{H}_\rm{dr}$ are strongly enhanced while the non-resonant processes are averaged out. In particular, we set $N_{ph}\left(t\right)=N_{ph,0}\left(1+\eta\cos\left(\omega_{m}t\right)\right)$ where $\eta\le 1$ is some constant modulation amplitude.  If we assume that there is no external potential besides the time-independent trapping potential, we find from equation (\ref{eq:potential})
\begin{equation}
	V_\rm{osc} - \delta\mu_\rm{osc} = \hbar\frac{g_0^2}{\Delta_A}\left(f_\rm{cav}^2 - \frac{1}{2}\right)N_{ph,0}\,\eta\cos\left(\omega_{m}t\right) 
\end{equation}
Through the rotating wave approximation (RWA), we discard terms that must oscillate at a non-zero frequency and cannot be brought into resonance. Then, to first and second order in the quasiparticle mode operators, we obtain
\begin{equation}
    \begin{split}
        \hat{H}_\rm{dr} & = \sum_n \left(\bar{P}_{n}\hat{b}_{n}e^{-i\left(\omega_{n}-\omega_{m}\right)t} + \bar{P}_{n}^*\hat{b}_{n}^{\dagger}e^{i\left(\omega_{n}-\omega_{m}\right)t}\right)\\
        & \hphantom{==} +\sum_{n} \left(\left(\bar{N}_{n}\hat{b}_{n}^{\dagger}\right)^{2}e^{i\left(2\omega_{n}-\omega_{m}\right)t}+\bar{N}_{n}^*\hat{b}_{n}^{2}e^{-i\left(2\omega_{n}-\omega_{m}\right)t}\right)\\
        & \hphantom{==}+\sum_{n,l<n}\left( \bar{M}_{nl}\hat{b}_{n}^{\dagger}\hat{b}_{l}e^{i\left(\omega_{n}-\omega_{l}-\omega_{m}\right)t} + \bar{M}_{nl}^*\hat{b}_{l}^{\dagger}\hat{b}_{n}e^{-i\left(\omega_{n}-\omega_{l}-\omega_{m}\right)t}\right)\\
        & \hphantom{==}+\sum_{n,l<n}\left(\bar{L}_{nl}\hat{b}_{n}\hat{b}_{l}e^{-i\left(\omega_{n}+\omega_{l}-\omega_{m}\right)t} + \bar{L}_{nl}^*\hat{b}_{n}^{\dagger}\hat{b}_{l}^{\dagger}e^{i\left(\omega_{n}+\omega_{l}-\omega_{m}\right)t}\right)
    \end{split}
    \label{eq:interaction_hamiltonian_resonantdriving}
\end{equation}
where the coefficients are given by equation (\ref{eq:coefficients}) with the replacement $V_\rm{osc} - \delta\mu_\rm{osc}\rightarrow \hbar g_0^2 (f_\rm{cav}^2 - 1/2)N_{ph,0}\,\eta/(2\Delta_A)$.

From the interaction Hamiltonian (\ref{eq:interaction_hamiltonian_resonantdriving}), we can see that we are able to to selectively drive particular quasiparticle interactions via the choice of the cavity field intensity oscillation frequency $\omega_m$. To be able to apply the RWA, we must ensure that the quasiparticle time scale (given by the inverse of the mode frequencies involved) and the measurement time are well separated.

\subsection{Beam splitting and mode swapping}
\label{subsec:beamsplit}

The first type of resonant interaction that we consider here is achieved with the resonance condition $\omega_m=\omega_n-\omega_l$, for which the interaction Hamiltonian (\ref{eq:interaction_hamiltonian_resonantdriving}) further reduces with the RWA to $\hat{H}_\rm{dr}=\bar{M}_{nl}\hat{b}_{n}^{\dagger}\hat{b}_{l} + \bar{M}_{nl}^*\hat{b}_{l}^{\dagger}\hat{b}_{n}$, assuming no other resonance conditions are met. The time evolution operator due to this Hamiltonian is thus
\begin{equation} \label{eq:mirror}
    \hat{U}_\rm{dr}=\exp\left[i\left(\mathcal{M}_{nl}\hat{b}_{n}^{\dagger}\hat{b}_{l} + \mathcal{M}_{nl}^*\hat{b}_{l}^{\dagger}\hat{b}_{n}\right)\right]
\end{equation}
where $\mathcal{M}_{nl}=\bar{M}_{nl}t/\hbar$. This has the form of a beam splitting or mode mixing operator, where the phase can be tuned both by the cavity intensity and interaction time. Calculating the evolution of the quasiparticle modes due to such an operation as $\hat{b}'=\hat{U}_\rm{dr}\hat{b}\hat{U}_\rm{dr}^\dagger$, we see that the effect can be written as
\begin{equation}
    \begin{pmatrix}
        \hat{b}'_{n}\\
        \hat{b}'_{l}
    \end{pmatrix}=
    \begin{pmatrix}
        \cos\left(|\mathcal{M}_{nl}|\right) & -ie^{i\theta_{nl}}\sin\left(|\mathcal{M}_{nl}|\right)\\
        -ie^{-i\theta_{nl}}\sin\left(|\mathcal{M}_{nl}|\right) & \cos\left(|\mathcal{M}_{nl}|\right)
    \end{pmatrix}
    \begin{pmatrix}
        \hat{b}_{n}\\
        \hat{b}_{l}
    \end{pmatrix}
\end{equation}
where $\theta_{nl}$ is the phase of $\mathcal{M}_{nl}$ defined such that $\mathcal{M}_{nl}=|\mathcal{M}_{nl}|e^{i\theta_{nl}}$. If $|\mathcal{M}_{nl}|=\pi/4$, then we are applying a symmetrical beam splitting operation, where two different quasiparticle modes are mixed with each other. If $|\mathcal{M}_{nl}|=\pi/2$, then we have a "mirror" or "swap" operation, where the photon occupation of two modes are swapped. These operations are highly versatile, and could be used to create novel, currently unfeasible, states. A "mirror" type operation could directly populate high order quasiparticle modes in a targeted way, starting from an easily attainable initial thermal state. A beam splitting operation is particularly interesting for both quantum metrology and fundamental quantum mechanics applications, as this creates entanglement between two modes. Examples by constructing a quasiparticle Mach-Zehnder interferometer and incorporating it into a  measurement scheme will be given in section \ref{sec:applications}.

\subsection{Displacement and squeezing}
\label{subsec:squeezing}

The second resonance condition we consider here is $\omega_m=\omega_n$. This resonance results as above in a time evolution of the form
\begin{equation}\label{eq:Udisp}
    \hat{U}_\rm{dr}=\exp\left[i\left(\mathcal{P}_n\hat{b}_{n}+\mathcal{P}_n^*\hat{b}^\dagger_{n}\right)\right]
\end{equation}
where $\mathcal{P}_{n} =\bar{P}_{n} t/\hbar$. This is a linear displacement operator, which over the interaction time $t$ creates a single mode coherent quasiparticle state in the mode labelled by $n$, with an average quasiparticle population of $|\mathcal{P}_{n}|^2$.

If instead we consider $\omega_m=2\omega_n$, the only resonant term results in a time evolution operator
\begin{equation}
    \hat{U}_\rm{dr}=\exp\left[\frac{i}{2}\left(\mathcal{N}_{n}\left(\hat{b}_{n}^{\dagger}\right)^{2} + \mathcal{N}_{n}^*\hat{b}_{n}^{2}\right)\right]
\end{equation}
where $\mathcal{N}_{n}=2\bar{N}_{n}t/\hbar$, which is a single mode squeezing operator where $\mathcal{N}_{n}$ plays the role of the squeezing parameter.

Finally, when $\omega_m=\omega_n+\omega_l$, we have a two mode squeezing operator
\begin{equation}
    \hat{U}_\rm{dr}=\exp\left[\frac{i}{2}\left(\mathcal{L}_{nl}\hat{b}_{n}\hat{b}_{l} + \mathcal{L}_{nl}^*\hat{b}_{n}^{\dagger}\hat{b}_{l}^{\dagger}\right)\right]
\end{equation}
where the squeezing parameter is given by $\mathcal{L}_{nl}=2\bar{L}_{nl}t/\hbar$.

These operations can in particular be used to generate coherent excitations of quasiparticles that may be used, for example, for sensing applications. In the next section, we will describe a specific sensing scheme; a quasiparticle Mach-Zehnder interferometer in frequency space.

\section{A quasiparticle Mach-Zehnder interferometer}
\label{sec:MZI}

The operations on the quasiparticle modes that we discussed in the previous section can be combined to realize quantum protocols with quasiparticles. As a specific example, we will discuss a quasiparticle Mach-Zehnder interferometer in this section. In particular, it can be used for sensing applications as we will explain later in our first example application; measuring the s-wave scattering length.

A Mach-Zehnder interferometer consists of two consecutive beam splitting operations acting on two modes, with a period of free evolution between them in which a phase difference $\Theta$ is accumulated. With the second beam splitting operation, the phase difference is imprinted on the individual modes by constructive or destructive interference, and can in principle be extracted. Here, we realize the beam splitting operations with the driving Hamiltonian as described in Sec. \ref{subsec:beamsplit} with $|\mathcal{M}_{nl}|=\pi/4$. With a vacuum state and a coherent state with the displacement parameter $\alpha_n$ as respective input states, it can be shown that the fundamental precision limit for phase estimation is given by \cite{demkowicz2015quantum}
\begin{equation}\label{eq:MZIsens}
	\Delta \Theta \ge \frac{1}{|\alpha_n|}\,,
\end{equation}
and is reached for $\Theta=\pi/2$. This corresponds to the standard quantum limit. The creation of the initial state can be realized in our setup using the displacement part of the driving Hamiltonian. Additional enhancement can be achieved, for example, with squeezed probe states which can be created with the squeezing operation discussed above. However, this quantum enhancement is strongly limited due to noise as we discuss in Sec. \ref{sec:damping}. Therefore, we do not consider quantum enhanced sensing in this article.

\section{Pulsed readout}
\label{sec:pulsed}

To apply the above-described manipulation methods to tasks in fundamental research and metrology, it must be possible to read out the state of the condensate. Several different techniques have been developed for measuring quasiparticle excitations of condensates. Excitations can be measured directly by measuring phase oscillations through "heterodyne detection" \cite{Katz:2004high} (see also \cite{Tozzo:2004pho}) or density perturbations by phase-contrast imaging \cite{Andrews:1996dir,Stamper:1998col}. Quasiparticle momenta can also be mapped onto internal states of atoms via doubly detuned Raman pulses \cite{Schuetzhold2006det}. Another option is time-of-flight measurements where the quasiparticle momentum is mapped onto free particle momenta after trap release and the propagation of the atoms falling in the gravitational field of the earth is measured (e.g. \cite{stamper1999excitation}). A particular version of this is to first split the elongated condensate in two parts that are then later brought into interference \cite{schumm2005matter,schweigler2017experimental,van2018projective,rauer2018recurrences,gluza2020quantum}.

In this section, we propose a method for imprinting the displacement of a single quasiparticle mode onto the cavity field phase, which can be read out with high precision through a homodyne detection scheme on the light leaving the cavity. The method is based on the approach of pulsed optomechanics presented, for example, in \cite{vanner2011pulsed,pikovski:thesis}. To this end, we assume that the dynamics of the quasiparticles is much slower than the measurement. This requires the measurement to be constructed from pulses shorter than the time scale of the quasiparticle dynamics. As we will show below, for typical experimental parameters, this requires light pulses with a length below the microsecond level to access high order quasiparticle modes, which is well within the capabilities of current technology. 

For the pulsed optomechanics scheme, the external read-out laser is tuned on resonance with the cavity field such that $\Delta_c = 0$ and we obtain the sum of driving and back-action Hamiltonian to leading order in $1/\sqrt{N_0}$
\begin{align}
       \hat{H}_\rm{plsd} =  \hat{H}_\rm{dr} + \hat{H}_\rm{ba} & = \sum_n  \kappa_{n} \left(\sqrt{N_{ph}} + \delta\hat{a}^\dagger + \delta\hat{a} \right) \left(\hat{b}_{n}e^{-i(\omega_{n}t_m - \theta_n)}  + \hat{b}_{n}^{\dagger}e^{i(\omega_{n}t_m - \theta_n)}\right)  \,.
\end{align}
If we consider the dynamics of the cavity field momentum quadrature perturbation $\hat{P}_L=i\left(\delta\hat{a}^\dagger-\delta\hat{a}\right)$, we find
\begin{equation}
    \begin{split}
        \hat{P}_{L}\left( \Delta t\right) & =\exp\left[\frac{i}{\hbar}\hat{H}_\rm{plsd} \Delta t\right]\hat{P}_{L}\exp\left[-\frac{i}{\hbar}\hat{H}_\rm{plsd} \Delta t\right]\\
        & =\hat{P}_{L} - \sum_n \frac{2\kappa_{n}}{\hbar} \left(\hat{b}_{n}e^{-i(\omega_{n}t_m - \theta_n)}  + \hat{b}_{n}^{\dagger}e^{i(\omega_{n}t_m - \theta_n)}\right) \Delta t \,.
    \end{split}
\end{equation}
Thus, we see that $\hat{P}_{L}$ depends on the quasiparticle displacement. Since we assume that the initial probe state of the light field is prepared such that $\langle \delta \hat{a} \rangle = 0$, $\langle \hat{a}^\dagger - \hat{a}\rangle = 0$ and $\langle \hat{a}^\dagger + \hat{a}\rangle = 2\sqrt{N_{ph}}\gg 1$, and the phase $\phi$ of a general coherent state is defined such that $\langle \hat{a}^\dagger + \hat{a}\rangle = 2\sqrt{N_{ph}} \cos \phi$ and $i\langle \hat{a}^\dagger - \hat{a}\rangle = 2\sqrt{N_{ph}} \sin \phi$, for small $\langle \hat{P}_{L}\left( \Delta t\right)\rangle$, we can interpret the shift of $\hat{P}_{L}$ as a small phase shift of the cavity mode state 
\begin{equation}
	\phi_\rm{ba} = \frac{\langle\hat{P}_{L}\left( \Delta t\right)\rangle}{2\sqrt{N_{ph}}} \,.
\end{equation}
This change of phase can, in principle, be read out by interfering the light from the cavity with a local phase reference, e.g. a homodyne measurement with fundamental precision limit given by the standard quantum limit
\begin{equation}
	\Delta \phi_\rm{ba} \ge \frac{1}{\sqrt{N_{ph}}}
\end{equation}
and signal to noise ratio
\begin{equation}\label{eq:SNR}
	\rm{SNR} = \frac{|\phi_\rm{ba}|}{\Delta \phi_\rm{ba}} \le \frac{1}{2} |\langle\hat{P}_{L}\left( \Delta t\right)\rangle|\,.
\end{equation}
A closer analysis of the measurement precision employing tools of quantum metrology reveals that the fundamental precision limit for the sensing of displacement of the quasiparticle modes saturates for large times and large coupling $\kappa_{2n_\rm{cav}}$; the detailed calculations can be found in Appendix \ref{app:fund-readout-limit}. To summarize, the fundamental precision limit depends on the Quantum Fisher Information (QFI), which quantifies the maximal amount of information that can be gained about a parameter encoded in a given state of the system from a measurement on the system (maximized over all possible measurements). For a coherent state of the cavity field with amplitude $\sqrt{N_{ph}}$ and quasiparticle mode with amplitude $\mathcal{P}_{n}$, the QFI for the estimation of $\mathcal{P}_{n}$ is
\begin{equation}
    F_{\rho_{F}}\left(\mathcal{P}_{n}\right)=\frac{16\chi^{2}}{1+4\chi^{2}}\,.
\end{equation}
From the QFI, we can then calculate the Cramer-Rao bound for the fundamental precision limit as
\begin{equation}
   (\Delta\mathcal{P}_{n})^2 \ge \frac{1}{F_{\rho_{F}}} =\frac{1}{4}+\frac{1}{16\chi^{2}}\rightarrow \frac{1}{4}  \quad\text{for}\quad \chi\rightarrow \infty\,,
\end{equation}
where $\chi = \kappa_{2n_\rm{cav}}\Delta t/\hbar$.

Note that, alternatively, the interaction described by $\hat{H}_\rm{ba}$ can be employed for mode cooling and state preparation by tuning the external driving laser frequency to the quasiparticle side bands in the sideband resolved regime, that is $\omega_n$ much larger than the cavity bandwidth. However, this cannot coexist with the state manipulation via intensity modulation and the pulsed optomechanics scheme which require the opposite regime, where $\omega_n$ is much smaller than the cavity bandwidth.

\section{Damping and decoherence}
\label{sec:damping}

For the above description of the condensate, we have considered the quasiparticle dynamics as lossless and we have neglected terms of third and fourth order in $\hat\vartheta$ in the expansion of $\hat{H}_\rm{total}$. In particular, terms of third order in $\hat\vartheta$ lead to Beliaev damping \cite{Giorgini:1998dam} and Landau damping \cite{Szepfalusy:1974on,Shi:1998fin,Fedichev:1998damp} which are significant loss mechanisms in condensates. Landau damping is induced by a process where a quasiparticle excitation scatters inelastically with a thermal quasiparticle excitation. Beliaev damping is the decay of a quasiparticle excitation (via the scattering with a condensate atom) into two quasiparticle excitations. While Landau damping can be suppressed by reducing the temperature, Beliaev damping is present also at zero temperature and scales very strongly with the quasi-particle momentum. For one-dimensional quasi-condensates, both processes are highly suppressed and fourth order processes become relevant (e.g. \cite{yuen2015enhanced}). For example for uniformity in the elongated direction, transverse harmonic trapping and low temperatures, the scattering-induced damping rate of high energy quasiparticles becomes (see e.g. $\Gamma^\rm{damp}$ on page 7 of \cite{mazets2011dynamics})
\begin{equation}\label{eq:1dloss}
	\gamma^\rm{1D}_\rm{sc} =  72 \sqrt{3} (\ln 4/3 )^2 \omega_\perp \left(\frac{\rho_\rm{1d} a_\rm{sc}^2 }{a_\perp}\right)^2  \,.
\end{equation}  
Another damping process that has to be taken into account is three-body losses, where three atoms interact. Two atoms form a molecule and the binding energy is transferred to the molecule and the third atom which are then ejected from the condensate. The corresponding decay constant is $\gamma_\rm{3B} := 3D\rho_0^2$, where $D$ is the three-body decay constant. For example, for $\rho_{\rm 0} = 10^{14}\,\rm{cm}^{-3}$ and a decay constant $D\sim 5.8\times 10^{-30}\,\rm{cm}^6\rm{s}^{-1}$ for rubidium atoms \cite{Burt:1997coh}, we find $\gamma_\rm{3B} = 3D\rho_{\rm c}^2 \sim 0.2 \,\rm{s}^{-1}$. Similarly, for an ytterbium BEC with $D \sim 4\times 10^{-30}\,\rm{cm}^6\rm{s}^{-1}$ \cite{Takuso:2003spin} and the same density, we obtain $\gamma_\rm{3B} \sim 0.1 \,\rm{s}^{-1}$. Three-body loss is also suppressed for one-dimensional quasi-condensates (see \cite{mehta_three-body_2007,haller_three-body_2011,tolra_observation_2004,haller_three-body_2011} and Appendix C of \cite{raetzel2021decay} for detailed discussion), so the effect of three-body loss can be limited by increasing the strengt of the transverse confinement of the condensate. 

In addition to the resulting loss of quasiparticle excitations, loss mechanisms are always accompanied by noise which leads to the decay of quantum enhancements in sensing applications \cite{Howl_2017,raetzel2021decay}. Both effects limit the time that coherent processes with quasiparticles can be performed in a condensate. To ensure that the approximations made for the description introduced in the previous section are valid for the examples below, we assume that all processes are performed on a time-scale much smaller than that of all loss mechanisms. We do not consider quantum enhancement explicitly in this article.

\section{Example applications}
\label{sec:applications}

Here, we present two examples of measurements that could be performed with the interactions we propose in this paper.

\subsection{Restriction to box traps}

For ease of calculation, we consider the trapping potential $V_0\left(z\right)$ in the elongated direction to be a uniform box potential of length $L$, e.g. as in \cite{rauer2018recurrences} or \cite{gaunt2013bose}, as the mode functions of the quasiparticle excitations have a particularly simple form. Furthermore, we assume that $L$ is much larger than the healing length of the condensate $\xi$, where $\xi = \hbar/\sqrt{2 m_a \tilde{g} \rho_\rm{0,1d}}$ and $\xi = \hbar/\sqrt{4 m_a \tilde{g} \rho_\rm{0,1d}}$ for three-dimensional BECs in transversely box-shaped traps and one-dimensional quasi-condensates in tight transverse harmonic traps, respectively, where $\rho_\rm{0,1d}=N_0/L$. This implies that the ground state wave function $\psi_0$ is almost constant over the length of the trap in the elongated direction besides a small region (on the length scale of $\xi$) at the boundary of the box potential where it quickly decays to zero \cite{Pitaevskii:2003bose} (see figure \ref{fig:ground-state} for an example). It follows from the stationary GP equation (\ref{eq:statgp}) that the chemical potential becomes $\mu=\tilde{g} \rho_\rm{0,1d}+V_0(z_0)$, where $z_0$ is chosen inside the box and $\overline{\delta V}$ is neglected as explained in Section \ref{sec:hamiltonian}.

To be able to derive analytical results, we restrict our considerations to two distinct regimes of quasiparticle modes. On the one hand, we consider modes where the atomic kinetic energy $\hbar^2 k_n^2/2m_a$ is much smaller than the interaction energy $\mu_0 = \tilde{g} \rho_\rm{0,1d}$, which implies that $k_n^2\xi^2\ll 1$ and therefore, the wavelength is much larger than the healing length. On the other hand, we consider high energy modes, where $\hbar^2 k_n^2/2m_a \gg \mu_0$, which implies that the wavelength is of the same order or much shorter than the healing length.

\subsubsection{Low energy modes}

For the low-energy modes, the ground state wave function appears box-like as $\psi_0=\chi_\rm{BT}/\sqrt{L}$, where $\chi_\rm{BT}$ is the characteristic function of the 1-dimensional box potential in the $z$-direction (i.e. $\chi_\rm{BT} = 1$ inside and $\chi_\rm{BT}=0$ outside the box, respectively). The abrupt decay of the density at the condensate's boundary leads to a delta-function-like term in the stationary BDG equations that govern the spatial dependence of the mode functions. This term implies Neumann boundary conditions on the quasiparticle modes. Furthermore, for the low-energy modes, we can apply the Thomas-Fermi approximation where the kinetic term in the stationary BDG equations are neglected. Therefore, the quasiparticle modes assume a particularly simple form 
\begin{equation}
    u_{n}^\rm{low}\left(z\right)=\alpha_n \psi_0(z) \varphi^\rm{c}_{n}\left(z\right)\text{ , and } \, \, v_{n}^\rm{low}\left(z\right)=\beta_n \psi_0(z)\varphi^\rm{c}_{n}\left(z\right),
\end{equation}
where
\begin{align}
	 \nonumber \alpha_n &= \left(\sigma_{n}^{-1}+\sigma_{n}\right)/2  \text{ , } \beta_n = \left(\sigma_{n}^{-1}-\sigma_{n}\right)/2\,\text{,}\\
    \sigma_{n}&= \left(1+2\tilde{g} \rho_\rm{0,1d}\left(\frac{\hbar^2k_n^2}{2m_a}\right)^{-1}\right)^{1/4} \text{ , and }\,\, \varphi^\rm{c}_{n}\left(z\right)= \sqrt{2 } \cos\left(k_{n}\left(z+\frac{L}{2}\right)\right)\text{ , }k_{n}=\frac{n\pi}{L}.
\end{align}
and we have chosen the boundaries of the trap potential at $z=-L/2$ and $z=L/2$. The mode frequencies become
\begin{equation}\label{eq:freq}
    \omega_{n}=\frac{\hbar k_{n}^{2}}{2m_a}\sigma_{n}^2 
\end{equation}
which simplifies to $\omega_{n}\approx c_{s}k_{n}$ where $c_{s}=\sqrt{\tilde{g}\rho_\rm{0,1d}/m_a}$ is the speed of sound in the low energy limit.

\subsubsection{High energy modes}

The high energy modes are defined by having kinetic energies much larger than the atom-atom interaction energy $g\rho_0$, which implies that these modes are dominated by their kinetic energy. Depending on the depth of the trapping potential, high energy quasiparticles can be free or bound. In the case of free propagation, we would recover the mode decomposition chosen in \cite{nagy2009nonlinear,ritsch2013cold}. In that case, quasiparticles would be lost eventually represented by atoms leaving the condensate and the length of the condensate would limit the time for driving and sensing. In the following, we assume that the trapping potential is deep enough for atoms with high energy quasiparticle momenta to be bound states. For the momenta $\hbar k$ of photons in the optical range, the recoil momentum $p_\rm{rec}=2\hbar k$ corresponds to a kinetic energy $p_\rm{rec}^2/(2m)$ of the order of $2\pi \hbar \times 10^4\,$Hz for $\rm{Rb}^{87}$-atoms. Trap depths of this order are consistently achieved with harmonic traps and are achievable in principle for box traps as well.

In practice, the box trap will not have perfectly steep walls on the length scale defined by the wavelength of high-energy modes, which is also on the same order or below the healing length $\xi$ of the condensate. Therefore, there are no simple boundary conditions for high-energy modes in contrast to the case of low-energy modes. The mode functions and energies have to be calculated numerically for explicit experimental situations. In Appendix \ref{app:toymodel}, we present a toy model with a trapezoid well where this is realized.

Assuming that $\xi\ll L$, in most parts of the trap the high-energy modes can be approximated as a linear combination of sine and cosine functions \footnote{Note that the conventional mode decomposition for uniform condensates would be modes of the form $\exp(i \hbar k_n z)$ and their complex conjugate. However, since the light mode is confined in a cavity, the light-atom interaction is symmetric under the exchange of the propagation direction of the quasiparticles and modes of the form $\cos(k_{n}(z+L/2))$ and $\sin(k_{n}(z+L/2))$ are directly addressed. The explicit coupling strength will depend on the position of the condensate inside the optical resonator.}, that is
\begin{eqnarray}
    u_{n}^\rm{high}\left(z\right)&\approx& \frac{1}{\sqrt{L}}\left( A^\rm{c}_n \varphi^\rm{c}_{n}\left(z\right) + A^\rm{s}_n \varphi^\rm{s}_{n}\left(z\right)\right)
\end{eqnarray}
where $\varphi^\rm{c}_{n}$ was defined above and 
\begin{align}
	\varphi^\rm{s}_{n}\left(z\right)= \sqrt{2 } \sin\left(k_{n}\left(z+\frac{L}{2}\right)\right)\,.
\end{align}
For the sake of simplicity and since we are only interested in principle limits of our scheme here, we assume in the following that the trap potential is optimized such that the contribution $A^\rm{s}_{n}$ is negligible for all modes that we couple to or address below (see Appendix \ref{app:toymodel} for an example) and set
\begin{eqnarray}
    u_{n}^\rm{high}\left(z\right)&\approx& \frac{1}{\sqrt{L}} \varphi^\rm{c}_{n}\left(z\right) \,
\end{eqnarray}
inside the trap and $u_{n}^\rm{high}=0$ outside the trap.

\subsubsection{The back-action Hamiltonian and pulsed readout}

By choosing the cavity mode appropriately (e.g. by choosing the length of the cavity, positioning of the cavity mirrors with respect to the atomic cloud), we can achieve that the coupling $\kappa_n$ is strongly emphasized for a specific mode and $\hat{H}_\rm{ba}$ in equation (\ref{eq:backaction_hamiltonian}) reduces approximately to the standard linearized optomechanical coupling Hamiltonian. 

In the following, we choose 
\begin{equation}
	f_\rm{cav}(z) = \sin\left(k_\rm{cav}\left(z+\frac{L}{2}\right)\right)
\end{equation}
and we assume that $k_\rm{cav}=n_\rm{cav} \pi/L$ where $n_\rm{cav}\in \mathbb{N}$. Dirichlet boundary conditions are then fulfilled at $z_{j_1}=-(j_1+1/2)L$ and $z_{j_2}=(j_2 + 1/2) L $ with $j_1,j_2 \in \mathbb{N}$, which implies that the cavity mirrors could be placed at any of these positions to realize the modes $f_\rm{cav}(z)$. The effective cavity mode volume becomes $\mathcal{V}_c=L_c \mathcal{A}_c/2$.

Based on the approximation made above for the mode functions, we find $\theta_n=0$ (as defined in equation     (\ref{eq:coupling})) and
\begin{eqnarray}
	\kappa_n &\approx & - \frac{\hbar g_0^2\sqrt{N_0 N_{ph}} }{2\sqrt{2}\Delta_A }\, \sigma_{2n_\rm{cav}}^{-1} \delta^{2n_\rm{cav}}_n \,, 
\end{eqnarray}
where $\delta^a_b$ is the Kronecker delta.
Then, we obtain that the coupling is only significant for $n = 2n_\rm{cav} $, that is $k_n = 2k_\rm{cav}$, which implies that momentum is conserved as in uniform condensates \cite{nagy2009nonlinear}.
The back-action Hamiltonian becomes
\begin{align}
        \hat{H}_\rm{plsd} & = \kappa_{2n_\rm{cav}} \left(\sqrt{N_{ph}} + \delta\hat{a}^\dagger + \delta\hat{a} \right) \left(\hat{b}_{2n_\rm{cav}}  + \hat{b}_{2n_\rm{cav}}^{\dagger} \right) 
    \label{eq:backaction_hamiltonian_2c}
\end{align}
where we have assumed that the measurement time $t_m$ is chosen such that $\omega_{2n_\rm{cav}} t_m$ is a multiple of $2\pi$. The momentum quadrature evolves accordingly as
\begin{equation}\label{eq:quadev}
        \hat{P}_{L}\left(t\right) = \hat{P}_{L}- \frac{2\kappa_{2n_\rm{cav}}}{\hbar} \left(\hat{b}_{2n_\rm{cav}} + \hat{b}_{2n_\rm{cav}}^{\dagger}\right) \Delta t \,
\end{equation}
which implies that the resulting phase shift can be associated with one specific quasiparticle mode.

\subsubsection{The driving Hamiltonian}

As for the back-action Hamiltonian, momentum conservation restricts the modes that the driving Hamiltonian allows access to. Based on the approximation made above for the mode functions, for the first order term in the driving Hamiltonian, we immediately find
\begin{align}\label{eq:Pn}
	\bar{P}_{n} &  \approx  -\frac{\eta}{2 } \sigma_n^{-1}\bar{\kappa} \sqrt{2 N_0} \,\delta^{2n_\rm{cav}}_n  \quad\text{where}\quad \bar{\kappa} = \hbar g_0^2 N_{ph,0}/(4\Delta_A)\,.
\end{align}
If a high-energy mode is involved, we find that all squeezing processes are suppressed and we do not write the expressions here. For low-energy modes, we obtain
\begin{equation}
	\bar{N}_{n} \approx  - \frac{\eta}{8}\left(\sigma_{n}^{-2}-\sigma_{n}^{2}\right) \bar{\kappa} \,\delta_{n}^{n_\rm{cav}} \,,
\end{equation} 
while the coefficient for two mode squeezing becomes
\begin{align}\label{eq:coefficients3}
    \bar{L}_{nl} & \approx  -\frac{\eta}{4}\left(\sigma_{n}^{-1}\sigma_{l}^{-1} - \sigma_{n}\sigma_{l}\right)  \bar{\kappa}   \, \left(\delta_{n-l}^{2n_\rm{cav}}  + \delta_{n+l}^{2n_\rm{cav}}\right)    \,.
\end{align}
The general expression for the coefficient for the beam-splitting operation is
\begin{align}\label{eq:coefficients2}
	      \bar{M}_{nl} & \approx   -\frac{\eta}{4}\left(\sigma_{n}^{-1}\sigma_{l}^{-1} + \sigma_{n}\sigma_{l}\right)  \bar{\kappa} \,\left(\delta_{n-l}^{2n_\rm{cav}}  + \delta_{n+l}^{2n_\rm{cav}}\right) \,.
\end{align}
where we took into account that $l<n$. The above list of coefficients will be the basis for explicit numerical examples in the next sections. 

In the expressions for $\bar{L}_{nl}$ and $\bar{M}_{nl}$, several options for combinations of momenta of coupled modes appear. For equidistant energies, each expression contains a momentum combination that, together with the energy condition of resonant driving, leads to resonant direct driving of a third mode. For example, for $\bar{L}_{nl}$, the condition $n+l=2n_\rm{cav}$ on the momenta and $\omega_n+\omega_l=\omega_m$ would lead to direct driving of the mode $n+l$. For $\bar{M}_{nl}$, the condition $n-l=2n_\rm{cav}$ on the momenta and $\omega_n-\omega_l=\omega_m$ would lead to direct driving of the mode $n-l$.  Since the coupling through $\bar{L}_{nl}$ and $\bar{M}_{nl}$ is weaker than the direct driving via $\bar{P}_{n}$ by a factor $2\sqrt{N_0}$, the direct driving of the additional modes would be significant, in general, even too large for the system to stay in the Bogoliubov regime. Therefore, the momentum combinations that lead to the driving of additional modes have to be avoided or at least one of the involved modes has to be a high-energy mode as the energy grows quadratically with the wave number in the high-energy regime.

\subsection{Explicit numerical examples}

In the following, we give two explicit examples that may be realized with typical parameters of state-of-the-art experimental systems.

The wavelength of typical lasers is below one micron and very high densities of the condensate would be necessary to obtain a healing length which is at least one order of magnitude smaller than the wavelength. This would limit the condensate lifetime and the quasiparticle coherence significantly (see section \ref{sec:damping}). Therefore, we will consider the first-order coupling (back-action and displacement) only for high-energy modes, i.e. for modes for which $\hbar^2k_n^2/(2m_a) \gg \tilde{g} \rho_\rm{0,1d}$. In all of the following examples, we consider two consecutive beam splitting or mode swapping operations that couple the high-energy mode to a low-energy mode.

\begin{figure}[b]
\centering
\includegraphics[width=10cm,angle=0]{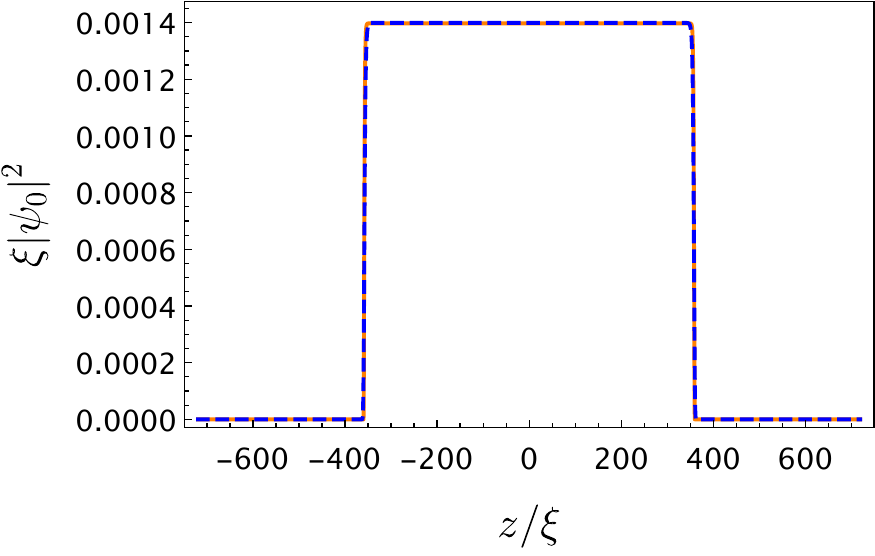}
\caption{\label{fig:ground-state} {\small  Absolute square $|\psi_0|^2$ of the ground state of a one-dimensional quasi-condensate for the parameters given in the text for the case of a box potential of depth $V_{0,\rm{b}}=80\mu_0\sim 2\hbar^2k_{n_\rm{high}}^2/(2m_a)$ (orange curve) and a more realistic continuous potential with finitely steep walls described by the function $V_0=-V_{0,\rm{b}} (\tanh(a(1 - (z/(\tilde{L}/2))^2)/4)-1)$ (blue dashed curve), where $a=200$, $\mu_0 = g_\rm{1d}\rho_\rm{0,1d}$ and $\tilde{L}=L/\sqrt{1 - (4\,\rm{arctanh}(1 - \mu_0/V_{0,\rm{b}})/a}$ (chosen such that $V_0=\mu_0$ at $z=-L/2$) . The gradient of the continuous potential is $\mp a V_{0,\rm{b}}/L$ at $\pm L/2$ which corresponds to an decrease/increase to half the box depth (the kinetic energy of the high-energy mode in the example) on a length scale $\sim L/100\sim 2\,\rm{\mu m}$. The two curves are practically indistinguishable 
on the length scale of wavelengths of low energy modes (roughly the length scale of the plot). See Appendix \ref{app:numericalGP} for a description of the numerical treatment. }}
\end{figure}

We consider a rubidium-87 1d quasi-condensate of $N_0=10^3$ atoms in a $200\mu$m long box trap at a density of $10^{14}$ atoms per cm$^3$, for which the healing length is approximately $\xi=2.77\times10^{-7}$m and the effective cross sectional area $\mathcal{A}$ whose value is equivalent to that of a circle with a radius  $a_\perp \sim 130$\,nm which could be created with a harmonic trap in the transverse direction with trap frequency $\sim 7\,$kHz. The scattering length of rubidium-87 is $\sim 5.18$\,nm \cite{Egorov:2013meas} and its atomic mass is $1.44\times 10^{-25}\,\rm{kg}$ which implies $\rho_\rm{1d}a_\rm{sc}\sim 0.03\ll 1 $ justifying the description of the quasi-condensate based on $\hat{H}_\rm{total}$. In figure \ref{fig:ground-state}, we show a plot of the absolute value squared of the ground state wave function in a proper box potential and a more realistic continuous potential approximating a box potential. We see that the change of the ground state is small and we consider the case of a discontinuous box potential for our analytical estimates below.

Rubidium atoms have two D-lines that can be used for the dispersive coupling; the D1-line is at $\nu_{D1}\sim 377$\,THz and the D2-line at $\nu_{D2}\sim 384$\,THz with a natural linewidth $\Gamma \sim 2\pi \times 6$\,Mhz and dipole moments of $d_1 \sim 2.5\times 10^{-29}$\,Cm and $d_2 \sim 3.6\times 10^{-29}$\,Cm for the D1 and D2 line, respectively. We will consider a first laser mode used for the preparation of the probe state through the displacement term in the driving Hamiltonian and choose a detuning of $\Delta_{A,1}/2\pi \sim +300$\,GHz from the D2-line. This implies that a mode with mode number $n_\rm{high} \sim 1020$ is addressed by direct driving. This mode is at a frequency of about $15$\,kHz for which $\hbar^2k_{n_\rm{high}}^2/(2m_a g_\rm{1d} \rho_\rm{0,1d}) \sim 40$ and the mode is in the high-energy regime. For the pulsed read-out, we consider a second laser mode with a much smaller detuning for $\Delta_{A,1'}/2\pi \sim 1$\,GHz to increase the coupling strength. This laser couples to the same quasiparticle mode as the driving laser. This is because the difference in the detuning is small enough such that the photon momenta do not differ significantly.

We consider two options for a laser mode for beam splitting/mode swapping operations. For our first example, a low-energy phonon mode with high frequency is advantageous. For the above parameters, the condition $\hbar^2k_n^2/(2m_a) \ll \tilde{g} \rho_\rm{0,1d}$ for low-energy modes limits the mode number to $n_\rm{low}\sim 50$ with a frequency of $\sim 170\,$Hz. To couple this mode to the high-energy mode necessitates a detuning of the beam splitting laser of $\Delta_{A,2}/2\pi \sim -20\,$THz from the D1-line. This detuning is larger than the frequency difference between the two lines which implies that both lines contribute to the driving process. For our second example application, a low frequency of the low-energy mode is advantageous. The mode number is already limited from below by the necessity to perform the driving over several periods of the mode while keeping the total driving time significantly below $1/\gamma^\rm{1D}_\rm{sc}$. Therefore, we choose a detuning of $\Delta_{A,2'}/2\pi \sim -300\,$GHz from the D1-line which implies a beam splitter/mode swapper coupling of the mode $n_\rm{high}\sim 1020$ to the mode $n_\rm{low}'\sim 20$ with a frequency of $\sim 70\,$Hz.

Assuming a cavity length of $L_\rm{cav} \sim 10$\,cm, the free spectral range becomes $\sim 3$\,GHz and it will be easy to find matching cavity modes in the chosen frequency ranges. Furthermore, we assume an effective cross sectional area of the cavity mode of 1\,mm$^2$. This implies single photon Rabi frequencies of $g_{0,2} \sim 180$\,kHz and $g_{0,1} \sim 130$\,kHz for the D2-line and D1-line respectively. 
To exclude all other processes besides the wanted beam splitting/mode swapping operation, we use a modulation of the intensity of the beam splitting laser mode in the cavity on resonance with the frequency difference between the high-energy and the low-energy mode and assume that each beam splitting/mode swapping operation lasts for $t_\rm{BS}=200$\,ms. Furthermore, we consider $100$\,ms for the accumulation of the signal between the beam splitting/mode swapping operations which gives a total time of the whole scheme of $500$\,ms plus the time for the preparation of the probe state, where we assumed that the read-out happens on a much shorter time-scale. For the parameters considered here, we find that $\gamma_\rm{sc}^\rm{1D} \sim 0.5$\,s$^{-1}$ for the mode $n_\rm{high}$ and we will neglect the decay of quasiparticles in the following.

With the above parameters, for the pulsed optomechanical read-out scheme, we obtain single-quasiparticle precision (SNR $\sim 1$ for $\langle b_{n_\rm{high}}\rangle = 1$, compare Eq. (\ref{eq:SNR}) with Eq. (\ref{eq:quadev})) if the accumulated total duration of the measurement is $\Delta t \gtrsim \hbar/(2 |\kappa_{2n_\rm{cav}}|) = \sqrt{2} \Delta_{A,1'}/(g_{0,2}^2\sqrt{N_0 N_{ph}})\sim 800\,\rm{ns}$ assuming $N_{ph}\sim 10^8$ photons in the cavity mode (corresponding to an intra-cavity power of $\sim 40$\,mW and an intensity of $4\,\rm{W/cm^2}$. ). 

To generate, for example, a coherent state of $N_{n_\rm{high}} = 10$ quasiparticles in the high-energy mode by resonant modulation of the intra-cavity power of the first laser mode for a duration of $t_\rm{cr}\sim 30$\,ms, an average number of photons in the cavity  $N_{ph,0}=|8\Delta_{A,1} \sqrt{N_{n_\rm{high}}}/(\eta g_{0,2}^2 t_\rm{cr} \sqrt{2N_0}) | \sim 10^3$ \footnote{Compare Eq. (\ref{eq:Pn}) with Eq. (\ref{eq:Udisp}).} (corresponding to $\sim 0.4\,\rm{\mu W}$ intra-cavity power) is sufficient given a modulation amplitude $\eta=1$ \footnote{ Note that we are interested in keeping the photon number sufficiently large to reduce the potential effects of photon number fluctuations (shot noise). The photon number can be increased, for example, by increasing the detuning. For sufficiently high photon numbers, the photon number fluctuations will average out due to the short lifetime of photons in the cavity of much less than the read-out time of 400\,ns.} . Furthermore, we obtain that the spatial maximum of the corresponding time-averaged light potential is $\rm{max}(\overline{\delta V})=8\hbar \sqrt{N_{n_\rm{high}}}/(t_\rm{cr}\eta \sqrt{2N_0}) \sim 0.008 \mu_0$, which justifies our treatment of the light potential as a small perturbation.

For the two driving laser modes employed for beam splitting/mode swapping with detunings $\Delta_{A,2}$ and $\Delta_{A,2'}$, we find a necessary photon number for a full mode swapping operation (i.e. $\bar{M}_{n_\rm{low}n_\rm{high}}t/\hbar \sim \pi/2$) at $t_\rm{BS}\sim 100$\,ms of $N_{ph,0}\gtrsim |4\pi(g_{0,1}^2/\Delta_{A,2} + g_{0,2}^2/(2\pi(\nu_{D1}-\nu_{D2})+\Delta_{A,2}))^{-1}/(\eta \alpha_{n_\rm{low}} t_\rm{BS}) | \sim 10^5$ and $N_{ph,0}\gtrsim |4\pi \Delta_{A,2'}/(\eta\alpha_{n_\rm{low}'} g_{0,1}^2 t_\rm{BS})| \sim 4\times 10^3$, respectively, for the Bogoliubov coefficients $\alpha_{n_\rm{low}}\sim 1.3$ and $\alpha_{n_\rm{low}'}\sim 1.8$ and a modulation amplitude $\eta=1$. For a beam splitting operation, half the number of photons may be employed.
For a full mode swapping operation, the spatial maximum of the time-averaged light-potential becomes $\rm{max}(\overline{\delta V})=4\pi\hbar/(\eta\alpha_{n_\rm{low}}t_\rm{BS}) \sim 0.02 \mu_0 $ for both driving laser frequencies. This value is sufficiently low with respect to the chemical potential to treat the resulting change of the condensate's ground state as a small perturbation, justifying our approach.

\subsubsection{Measuring the s-wave scattering length}

A change of the s-wave scattering length from $a_{\rm{sc},1}$ to some $a_{\rm{sc},2}$ will change the frequency of low-energy modes through the dispersion relation (\ref{eq:freq}), i.e. $\omega_{n_\rm{low}} \propto \sqrt{a_\rm{sc}}$. The frequency of the high-energy mode can be approximated as unaffected. Therefore, the relevant phase shift accumulated between the two modes of the quasiparticle MZI is $\Delta \phi = \Delta\omega_{n_\rm{low}} t_\rm{int}$, where $\Delta\omega_{n_\rm{low}} =\omega_{n_\rm{low}}(a_{\rm{sc},1})-\omega_{n_\rm{low}}(a_{\rm{sc},2})$ is the frequency shift due to the change of the scattering length and $t_\rm{int}$ is the time during which the s-wave scattering length is modified. We chose $t_\rm{int}=100$\,ms, equivalent to the time between the two beam splitting operations of the MZI. By Gaussian error propagation, we obtain for the fundamental relative precision limit for estimations of any small change in the scattering length from Eq. \ref{eq:MZIsens} as
\begin{equation}
	\frac{\Delta a_\rm{sc}}{a_\rm{sc}} \ge \left|\frac{d \Delta\omega_{n_\rm{low}}}{d a_\rm{sc}} t_\rm{int}\right|^{-1} \frac{1}{a_\rm{sc} \sqrt{N_\rm{high}}} \sim \frac{2}{\omega_{n_\rm{low}} t_\rm{int} \sqrt{N_{n_\rm{high}}}}
\end{equation}
where we have assumed that the low-energy mode is initially in the ground state and the high-energy mode is brought into a coherent probe state with quasiparticle number $N_{n_\rm{high}}$. If we assume that the probe state contains $N_{n_\rm{high}} = 10$ quasiparticles and use $\omega_{n_\rm{low}}\sim 2\pi\times 170\,$Hz for $n_\rm{low}\sim 50$, we find for the relative precision $\Delta a_\rm{sc}/a_\rm{sc}\gtrsim  0.006$. 

This value for the precision limit can be optimized by considering longer interaction times which is, however, limited by the decay of quasiparticles due to the damping mechanisms described above. A larger number of quasiparticles in the probe state would also be advantageous; this is however limited to be much smaller than the number of atoms in the quasi-condensate, which is limited due to the size of the trap and the necessity to keep the density low to avoid three-body losses.

\subsubsection{Sensing oscillating force gradients}
\label{subsec:oscillating_mass}

As a second example application, we consider a temporally oscillating force gradient represented by the potential $\delta V=-z^2 G_0 \sin\left(\Omega t\right)$. In \cite{ratzel_dynamical_2018}, it has been shown that the effect on quasiparticle modes is stronger for smaller frequencies and enhanced by resonance. Therefore, we assume that $\Omega$ is on resonance with the low-energy mode $n_\rm{low}$. In the Heisenberg picture and to leading order in the quasiparticle field $\hat\vartheta$, the interaction Hamiltonian takes the form (see Appendix \ref{sec:derivationintHamPi})
\begin{equation}\label{eq:intHamPi}
    \hat{H}_\rm{int}= \Pi_{n_\rm{low}}\left(\hat{b}_{n_\rm{low}} - \hat{b}_{n_\rm{low}}^{\dagger}\right),
\end{equation}
where, 
\begin{eqnarray}
	\Pi_{n_\rm{low}} &=& i\frac{\sqrt{N_0} G_0}{2 L \sigma_{n_\rm{low}}}   \int_{-L/2}^{L/2} dz\, \left(z^2 - \frac{L^2}{12} \right) \, \varphi_{n_\rm{low}}^\rm{c}(z)  = i\frac{\sqrt{2 N_0} G_0}{\sigma_{n_\rm{low}}k_{n_\rm{low}}^2} \nonumber \\
	&\approx & i  G_0  \sqrt{2\pi N_0 \rho_0 a_\rm{sc}}  \left(\frac{\hbar}{m_a \omega_n} \right)^{3/2}  \,,
\end{eqnarray}
for $n_\rm{low}$ even and $\Pi_{n_\rm{low}}=0$ for $n_\rm{low}$ odd, and through the RWA, we disregard terms that are off resonant. The effect of the oscillating force is to create quasiparticles through a linear displacement with amplitude $|\Pi_{n_\rm{low}} t/\hbar|$. 

We first apply a mode swapping operation between modes $n_\rm{low}=20$ and $n_\rm{high}=1020$, as detailed in section \ref{subsec:beamsplit}. This is advantageous as the low-energy mode has higher thermal occupancy than the high-energy mode we consider for read-out. In particular, we can assume that the probe state is the vacuum state and that the time-scale of thermalization which is bound from below by  $1/\gamma_\rm{sc}^\rm{1D}$ is much larger than the time scale of the full sensing protocol. We then let the quasiparticle modes evolve while the force gradient oscillates, creating a displaced state in mode $n_\rm{low}=20$. A second mode swapping operation is applied to bring the displacement signal back to $n_\rm{high}$, where it can be read out with an appropriate read-out scheme, for example, the pulsed scheme outlined in section \ref{sec:pulsed}.

We assume that the read-out achieves a single-quasiparticle precision in a single-shot experiment. Given the above parameters, one quasiparticle is created after $t_\rm{int}=100\,$ms for a minimal force gradient
\begin{equation}
	G_{0,\rm{min}} = \frac{\left(m_a \omega_{n_\rm{low}}\right)^{3/2}}{t_\rm{int} \sqrt{ 2\pi \hbar N_0 \rho_0 a_\rm{sc}}} \sim  10^{-23}\,\rm{Nm^{-1}} \,.
\end{equation}
This value corresponds to the force gradient induced by a harmonic potential with frequency $\sqrt{G_{0,\rm{min}}/m_a}/(2\pi) \sim 1$\,Hz acting on the atoms. Repeating the experiment for about a week, that is about $10^4$ times, would reduce the minimal force gradient by another 2 orders of magnitude to $\sqrt{G_{0,\rm{min}}/m_a}/(2\pi) \sim 10$\,mHz. This theoretical prediction can be compared with the experiment described in \cite{Obrecht:2007meas}, where a precision for a time-independent Casimir-Polder force gradient slightly below 100\,mHz has been reached by employing the center of mass motion of a BEC. However, one has to take into account that sources of noise and other systematic errors which play an important role in \cite{Obrecht:2007meas} are not considered here.

The minimal force gradient can, in principle, be decreased by choosing a lower frequency of the probe mode. For example, employing $n=2$ instead of $n=20$ would lead to a decrease of $G_{0,\rm{min}}$ by a factor $\sim 30$. However, to ensure that the mode swapping only couples to the low-energy mode, the driving time has to be increased by a factor 10 as well, which increases the total interaction time to the same value as the time scale of the decay of high-energy quasiparticles $1/\gamma_\rm{sc}^\rm{1D}$. As above, an increase of the number of atoms seems to be advantageous as it would increase the driving strength. However, it would also lead to a strong decrease in condensate lifetime which overcompensates the increase in driving strength.

\section{Conclusion and Discussion}
\label{sec:conclusion}

We developed a framework for the description of cavity optomechanics with trapped condensates to describe the manipulation and read-out of quasiparticle states with laser fields. The set of possible operations includes displacement, single-mode squeezing, two-mode squeezing and mode-mixing. In particular, displacement and mode-mixing can be used to create a quasiparticle Mach-Zehnder Interferometer (MZI) in frequency space. We have presented example applications and considered parameters of state-of-the-art technology to enable estimates of the fundamental limitations of the scheme for quantum metrology. For example, the measurement scheme for force gradients that we discussed may be employed to measure the thermal Casimir-Polder force as in \cite{Obrecht:2007meas} and the quasiparticle MZI may be employed for sensing effects that lead to a differential change of the frequencies of the quasiparticle modes. However, we found that the fundamental sensitivity bounds for state-of-the-art parameters are not very promising and significant technological progress would be needed to overcome the inherent limitations.  

The most severe restrictions arise from noise in the condensate due to quasiparticle-quasiparticle interactions and atom loss. Such effects limit the lifetime of the condensate, and thus limit the time for signal accumulation and state manipulation. We conclude that, in practice, displacements can only be induced in modes with particle-like dispersion due the short wavelengths of lasers which can readily be confined to cavities. For these high-energy modes, Beliaev damping becomes strongly pronounced in three-dimensional ultra-cold Bose gases. Therefore, we considered one-dimensional quasi-condensates in our example applications. It would be interesting to re-assess our example applications if laser-cavity systems with wavelengths in the range of $10-100\,\rm{\mu m}$ were available. In that case, low-energy quasiparticle modes could be addressed directly and three-dimensional quantum gases could be considered allowing for much higher numbers of condensate atoms and quasiparticles in the probe states. Another option to access low-lying phonon modes directly would be to arrange the beam line of the driving laser at a large angle to the longitudional axis of the BEC \footnote{In this case, momentum transfer to the phonons would couple photons out of the cavity mode which implies a different coupling Hamiltonian than the one used in this article.}, which however does not lift the restriction to one-dimensional condensates, as the mode structure has to be confined to restrict the coupling to longitudinal modes. 

We estimate the fundamental capabilities of our scheme here without considering experimental imperfections such as vibrations, trap instability, imperfect boundary conditions and atom losses. As these would depend on a specific implementation, we leave these calculations to future work, which could assist in further identifying the technological advancements necessary to implement our scheme.

It would be interesting to investigate similar sensing and mode manipulation schemes with superfluid helium which is much more stable, enables much higher coherence times and direct driving of modes with wavelengths in the microwave range (see for example \cite{Lorenzo_2014}).

\section*{acknowledgements}

We thank Matthias Sonnleitner, Ralf Sch\"utzhold, Mira Maiwöger, Naceur Gaaloul, Thomas Kiel, Philip Kristensen, Francesco Intravaia and Dan-Nha Huynh for helpful remarks and discussions. We thank Matthias Sonnleitner for generously providing the numerical code for the calculation of the condensate ground state. D.R. and D.H. acknowledge support by the Marie Skłodowska-Curie Action IF program -- Project-Name "Phononic Quantum Sensors for Gravity" (PhoQuS-G) -- Grant-Number 832250.

\bibliographystyle{IEEEtran}
\bibliography{phonon-BS}

\begin{thebibliography}{100}
\providecommand{\url}[1]{#1}
\csname url@samestyle\endcsname
\providecommand{\newblock}{\relax}
\providecommand{\bibinfo}[2]{#2}
\providecommand{\BIBentrySTDinterwordspacing}{\spaceskip=0pt\relax}
\providecommand{\BIBentryALTinterwordstretchfactor}{4}
\providecommand{\BIBentryALTinterwordspacing}{\spaceskip=\fontdimen2\font plus
\BIBentryALTinterwordstretchfactor\fontdimen3\font minus
  \fontdimen4\font\relax}
\providecommand{\BIBforeignlanguage}[2]{{%
\expandafter\ifx\csname l@#1\endcsname\relax
\typeout{** WARNING: IEEEtran.bst: No hyphenation pattern has been}%
\typeout{** loaded for the language `#1'. Using the pattern for}%
\typeout{** the default language instead.}%
\else
\language=\csname l@#1\endcsname
\fi
#2}}
\providecommand{\BIBdecl}{\relax}
\BIBdecl

\bibitem{ludlow2015optical}
\BIBentryALTinterwordspacing
A.~D. Ludlow, M.~M. Boyd, J.~Ye, E.~Peik, and P.~O. Schmidt, ``Optical atomic
  clocks,'' \emph{Rev. Mod. Phys.}, vol.~87, pp. 637--701, Jun 2015. [Online].
  Available: \url{https://link.aps.org/doi/10.1103/RevModPhys.87.637}
\BIBentrySTDinterwordspacing

\bibitem{lewenstein2007ultracold}
\BIBentryALTinterwordspacing
M.~Lewenstein, A.~Sanpera, V.~Ahufinger, B.~Damski, A.~Sen(De), and U.~Sen,
  ``Ultracold atomic gases in optical lattices: mimicking condensed matter
  physics and beyond,'' \emph{Advances in Physics}, vol.~56, no.~2, pp.
  243--379, 2007. [Online]. Available:
  \url{https://doi.org/10.1080/00018730701223200}
\BIBentrySTDinterwordspacing

\bibitem{Dalibard2011coll}
\BIBentryALTinterwordspacing
J.~Dalibard, F.~Gerbier, G.~Juzeli\ifmmode~\bar{u}\else \={u}\fi{}nas, and
  P.~\"Ohberg, ``Colloquium: Artificial gauge potentials for neutral atoms,''
  \emph{Rev. Mod. Phys.}, vol.~83, pp. 1523--1543, Nov 2011. [Online].
  Available: \url{https://link.aps.org/doi/10.1103/RevModPhys.83.1523}
\BIBentrySTDinterwordspacing

\bibitem{cooper2019topological}
\BIBentryALTinterwordspacing
N.~R. Cooper, J.~Dalibard, and I.~B. Spielman, ``Topological bands for
  ultracold atoms,'' \emph{Rev. Mod. Phys.}, vol.~91, p. 015005, Mar 2019.
  [Online]. Available:
  \url{https://link.aps.org/doi/10.1103/RevModPhys.91.015005}
\BIBentrySTDinterwordspacing

\bibitem{peters1999measurement}
\BIBentryALTinterwordspacing
A.~Peters, K.~Y. Chung, and S.~Chu, ``Measurement of gravitational acceleration
  by dropping atoms,'' \emph{Nature}, vol. 400, no. 6747, pp. 849--852, 1999.
  [Online]. Available: \url{https://doi.org/10.1038/23655}
\BIBentrySTDinterwordspacing

\bibitem{mcguirk2002sensitive}
\BIBentryALTinterwordspacing
J.~M. McGuirk, G.~T. Foster, J.~B. Fixler, M.~J. Snadden, and M.~A. Kasevich,
  ``Sensitive absolute-gravity gradiometry using atom interferometry,''
  \emph{Phys. Rev. A}, vol.~65, p. 033608, Feb 2002. [Online]. Available:
  \url{https://link.aps.org/doi/10.1103/PhysRevA.65.033608}
\BIBentrySTDinterwordspacing

\bibitem{geiger2011detecting}
\BIBentryALTinterwordspacing
R.~Geiger, V.~M{\'e}noret, G.~Stern, N.~Zahzam, P.~Cheinet, B.~Battelier,
  A.~Villing, F.~Moron, M.~Lours, Y.~Bidel \emph{et~al.}, ``Detecting inertial
  effects with airborne matter-wave interferometry,'' \emph{Nature
  communications}, vol.~2, no.~1, pp. 1--7, 2011. [Online]. Available:
  \url{https://doi.org/10.1038/ncomms1479}
\BIBentrySTDinterwordspacing

\bibitem{bidel2018absolute}
\BIBentryALTinterwordspacing
Y.~Bidel, N.~Zahzam, C.~Blanchard, A.~Bonnin, M.~Cadoret, A.~Bresson,
  D.~Rouxel, and M.~Lequentrec-Lalancette, ``Absolute marine gravimetry with
  matter-wave interferometry,'' \emph{Nature communications}, vol.~9, no.~1,
  pp. 1--9, 2018. [Online]. Available:
  \url{https://doi.org/10.1038/s41467-018-03040-2}
\BIBentrySTDinterwordspacing

\bibitem{rosi2014precision}
\BIBentryALTinterwordspacing
G.~Rosi, F.~Sorrentino, L.~Cacciapuoti, M.~Prevedelli, and G.~Tino, ``Precision
  measurement of the newtonian gravitational constant using cold atoms,''
  \emph{Nature}, vol. 510, no. 7506, pp. 518--521, 2014. [Online]. Available:
  \url{https://doi.org/10.1038/nature13433}
\BIBentrySTDinterwordspacing

\bibitem{parker2018measurement}
\BIBentryALTinterwordspacing
R.~H. Parker, C.~Yu, W.~Zhong, B.~Estey, and H.~M{\"u}ller, ``Measurement of
  the fine-structure constant as a test of the standard model,''
  \emph{Science}, vol. 360, no. 6385, pp. 191--195, 2018. [Online]. Available:
  \url{https://doi.org/10.1126/science.aap7706}
\BIBentrySTDinterwordspacing

\bibitem{Fray2004atomic}
\BIBentryALTinterwordspacing
S.~Fray, C.~A. Diez, T.~W. H\"ansch, and M.~Weitz, ``Atomic interferometer with
  amplitude gratings of light and its applications to atom based tests of the
  equivalence principle,'' \emph{Phys. Rev. Lett.}, vol.~93, p. 240404, Dec
  2004. [Online]. Available:
  \url{https://link.aps.org/doi/10.1103/PhysRevLett.93.240404}
\BIBentrySTDinterwordspacing

\bibitem{Schlippert2014quantum}
\BIBentryALTinterwordspacing
D.~Schlippert, J.~Hartwig, H.~Albers, L.~L. Richardson, C.~Schubert, A.~Roura,
  W.~P. Schleich, W.~Ertmer, and E.~M. Rasel, ``Quantum test of the
  universality of free fall,'' \emph{Phys. Rev. Lett.}, vol. 112, p. 203002,
  May 2014. [Online]. Available:
  \url{https://link.aps.org/doi/10.1103/PhysRevLett.112.203002}
\BIBentrySTDinterwordspacing

\bibitem{hamilton_atom-interferometry_2015}
\BIBentryALTinterwordspacing
P.~Hamilton, M.~Jaffe, P.~Haslinger, Q.~Simmons, H.~Müller, and J.~Khoury,
  ``\BIBforeignlanguage{en}{Atom-interferometry constraints on dark energy},''
  \emph{\BIBforeignlanguage{en}{Science}}, vol. 349, no. 6250, pp. 849--851,
  Aug. 2015. [Online]. Available:
  \url{https://science.sciencemag.org/content/349/6250/849}
\BIBentrySTDinterwordspacing

\bibitem{Muentinga:2013int}
\BIBentryALTinterwordspacing
H.~M\"untinga, H.~Ahlers, M.~Krutzik, A.~Wenzlawski, S.~Arnold, D.~Becker,
  K.~Bongs, H.~Dittus, H.~Duncker, N.~Gaaloul, C.~Gherasim, E.~Giese,
  C.~Grzeschik, T.~W. H\"ansch, O.~Hellmig, W.~Herr, S.~Herrmann, E.~Kajari,
  S.~Kleinert, C.~L\"ammerzahl, W.~Lewoczko-Adamczyk, J.~Malcolm, N.~Meyer,
  R.~Nolte, A.~Peters, M.~Popp, J.~Reichel, A.~Roura, J.~Rudolph,
  M.~Schiemangk, M.~Schneider, S.~T. Seidel, K.~Sengstock, V.~Tamma,
  T.~Valenzuela, A.~Vogel, R.~Walser, T.~Wendrich, P.~Windpassinger, W.~Zeller,
  T.~van Zoest, W.~Ertmer, W.~P. Schleich, and E.~M. Rasel, ``Interferometry
  with bose-einstein condensates in microgravity,'' \emph{Phys. Rev. Lett.},
  vol. 110, p. 093602, Feb 2013. [Online]. Available:
  \url{https://link.aps.org/doi/10.1103/PhysRevLett.110.093602}
\BIBentrySTDinterwordspacing

\bibitem{van2010bose}
\BIBentryALTinterwordspacing
T.~van Zoest, N.~Gaaloul, Y.~Singh, H.~Ahlers, W.~Herr, S.~Seidel, W.~Ertmer,
  E.~Rasel, M.~Eckart, E.~Kajari \emph{et~al.}, ``Bose-einstein condensation in
  microgravity,'' \emph{Science}, vol. 328, no. 5985, pp. 1540--1543, 2010.
  [Online]. Available: \url{https://doi.org/10.1126/science.1189164}
\BIBentrySTDinterwordspacing

\bibitem{gaaloul2014precision}
N.~Gaaloul, J.~Hartwig, C.~Schubert, W.~Ertmer, and E.~Rasel, ``Precision
  interferometry with bose-einstein condensates,'' in \emph{Atom
  interferometry}.\hskip 1em plus 0.5em minus 0.4em\relax IOS Press, 2014, pp.
  657--689.

\bibitem{Abend:2016atom}
\BIBentryALTinterwordspacing
S.~Abend, M.~Gebbe, M.~Gersemann, H.~Ahlers, H.~M\"untinga, E.~Giese,
  N.~Gaaloul, C.~Schubert, C.~L\"ammerzahl, W.~Ertmer, W.~P. Schleich, and
  E.~M. Rasel, ``Atom-chip fountain gravimeter,'' \emph{Phys. Rev. Lett.}, vol.
  117, p. 203003, Nov 2016. [Online]. Available:
  \url{https://link.aps.org/doi/10.1103/PhysRevLett.117.203003}
\BIBentrySTDinterwordspacing

\bibitem{Hardman:2016sim}
\BIBentryALTinterwordspacing
K.~S. Hardman, P.~J. Everitt, G.~D. McDonald, P.~Manju, P.~B. Wigley, M.~A.
  Sooriyabandara, C.~C.~N. Kuhn, J.~E. Debs, J.~D. Close, and N.~P. Robins,
  ``Simultaneous precision gravimetry and magnetic gradiometry with a
  bose-einstein condensate: A high precision, quantum sensor,'' \emph{Phys.
  Rev. Lett.}, vol. 117, p. 138501, Sep 2016. [Online]. Available:
  \url{https://link.aps.org/doi/10.1103/PhysRevLett.117.138501}
\BIBentrySTDinterwordspacing

\bibitem{Asenbaum:2017pha}
\BIBentryALTinterwordspacing
P.~Asenbaum, C.~Overstreet, T.~Kovachy, D.~D. Brown, J.~M. Hogan, and M.~A.
  Kasevich, ``Phase shift in an atom interferometer due to spacetime curvature
  across its wave function,'' \emph{Phys. Rev. Lett.}, vol. 118, p. 183602, May
  2017. [Online]. Available:
  \url{https://link.aps.org/doi/10.1103/PhysRevLett.118.183602}
\BIBentrySTDinterwordspacing

\bibitem{becker2018space}
\BIBentryALTinterwordspacing
D.~Becker, M.~D. Lachmann, S.~T. Seidel, H.~Ahlers, A.~N. Dinkelaker,
  J.~Grosse, O.~Hellmig, H.~M{\"u}ntinga, V.~Schkolnik, T.~Wendrich
  \emph{et~al.}, ``Space-borne bose--einstein condensation for precision
  interferometry,'' \emph{Nature}, vol. 562, no. 7727, pp. 391--395, 2018.
  [Online]. Available: \url{http://dx.doi.org/10.1038/s41586-018-0605-1}
\BIBentrySTDinterwordspacing

\bibitem{aveline2020observation}
\BIBentryALTinterwordspacing
D.~C. Aveline, J.~R. Williams, E.~R. Elliott, C.~Dutenhoffer, J.~R. Kellogg,
  J.~M. Kohel, N.~E. Lay, K.~Oudrhiri, R.~F. Shotwell, N.~Yu \emph{et~al.},
  ``Observation of bose--einstein condensates in an earth-orbiting research
  lab,'' \emph{Nature}, vol. 582, no. 7811, pp. 193--197, 2020. [Online].
  Available: \url{https://doi.org/10.1038/s41586-020-2346-1}
\BIBentrySTDinterwordspacing

\bibitem{bloch2012quantum}
\BIBentryALTinterwordspacing
I.~Bloch, J.~Dalibard, and S.~Nascimbene, ``Quantum simulations with ultracold
  quantum gases,'' \emph{Nature Physics}, vol.~8, no.~4, pp. 267--276, 2012.
  [Online]. Available: \url{https://doi.org/10.1038/nphys2259}
\BIBentrySTDinterwordspacing

\bibitem{gring2012relaxation}
\BIBentryALTinterwordspacing
M.~Gring, M.~Kuhnert, T.~Langen, T.~Kitagawa, B.~Rauer, M.~Schreitl, I.~Mazets,
  D.~A. Smith, E.~Demler, and J.~Schmiedmayer, ``Relaxation and
  prethermalization in an isolated quantum system,'' \emph{Science}, vol. 337,
  no. 6100, pp. 1318--1322, 2012. [Online]. Available:
  \url{https://doi.org/10.1126/science.1224953}
\BIBentrySTDinterwordspacing

\bibitem{rauer2018recurrences}
\BIBentryALTinterwordspacing
B.~Rauer, S.~Erne, T.~Schweigler, F.~Cataldini, M.~Tajik, and J.~Schmiedmayer,
  ``Recurrences in an isolated quantum many-body system,'' \emph{Science}, vol.
  360, no. 6386, pp. 307--310, 2018. [Online]. Available:
  \url{https://www.science.org/doi/abs/10.1126/science.aan7938}
\BIBentrySTDinterwordspacing

\bibitem{Michael2019from}
\BIBentryALTinterwordspacing
M.~H. Michael, J.~Schmiedmayer, and E.~Demler, ``From the moving piston to the
  dynamical casimir effect: Explorations with shaken condensates,'' \emph{Phys.
  Rev. A}, vol.~99, p. 053615, May 2019. [Online]. Available:
  \url{https://link.aps.org/doi/10.1103/PhysRevA.99.053615}
\BIBentrySTDinterwordspacing

\bibitem{barcelo2001analogue}
\BIBentryALTinterwordspacing
C.~Barcelo, S.~Liberati, and M.~Visser, ``Analogue gravity from bose-einstein
  condensates,'' \emph{Classical and Quantum Gravity}, vol.~18, no.~6, p. 1137,
  2001. [Online]. Available: \url{https://doi.org/10.1088/0264-9381/18/6/312}
\BIBentrySTDinterwordspacing

\bibitem{lahav2010realization}
\BIBentryALTinterwordspacing
O.~Lahav, A.~Itah, A.~Blumkin, C.~Gordon, S.~Rinott, A.~Zayats, and
  J.~Steinhauer, ``Realization of a sonic black hole analog in a bose-einstein
  condensate,'' \emph{Phys. Rev. Lett.}, vol. 105, p. 240401, Dec 2010.
  [Online]. Available:
  \url{https://link.aps.org/doi/10.1103/PhysRevLett.105.240401}
\BIBentrySTDinterwordspacing

\bibitem{steinhauer2014observation}
\BIBentryALTinterwordspacing
J.~Steinhauer, ``Observation of self-amplifying hawking radiation in an
  analogue black-hole laser,'' \emph{Nature Physics}, vol.~10, no.~11, pp.
  864--869, 2014. [Online]. Available: \url{https://doi.org/10.1038/nphys3104}
\BIBentrySTDinterwordspacing

\bibitem{fischer2004quantum}
\BIBentryALTinterwordspacing
U.~R. Fischer and R.~Sch\"utzhold, ``Quantum simulation of cosmic inflation in
  two-component bose-einstein condensates,'' \emph{Phys. Rev. A}, vol.~70, p.
  063615, Dec 2004. [Online]. Available:
  \url{https://link.aps.org/doi/10.1103/PhysRevA.70.063615}
\BIBentrySTDinterwordspacing

\bibitem{bravo2015analog}
\BIBentryALTinterwordspacing
T.~Bravo, C.~Sab{\'\i}n, and I.~Fuentes, ``Analog quantum simulation of
  gravitational waves in a bose-einstein condensate,'' \emph{EPJ Quantum
  Technology}, vol.~2, no.~1, pp. 1--9, 2015. [Online]. Available:
  \url{https://doi.org/10.1140/epjqt16}
\BIBentrySTDinterwordspacing

\bibitem{hartley2018analogue}
\BIBentryALTinterwordspacing
D.~Hartley, T.~Bravo, D.~R{\"a}tzel, R.~Howl, and I.~Fuentes, ``Analogue
  simulation of gravitational waves in a $3+1$-dimensional bose-einstein
  condensate,'' \emph{Phys. Rev. D}, vol.~98, p. 025011, Jul 2018. [Online].
  Available: \url{https://link.aps.org/doi/10.1103/PhysRevD.98.025011}
\BIBentrySTDinterwordspacing

\bibitem{Obrecht:2007meas}
\BIBentryALTinterwordspacing
J.~M. Obrecht, R.~J. Wild, M.~Antezza, L.~P. Pitaevskii, S.~Stringari, and
  E.~A. Cornell, ``Measurement of the temperature dependence of the
  casimir-polder force,'' \emph{Phys. Rev. Lett.}, vol.~98, p. 063201, Feb
  2007. [Online]. Available:
  \url{https://link.aps.org/doi/10.1103/PhysRevLett.98.063201}
\BIBentrySTDinterwordspacing

\bibitem{Antezza:2004effect}
\BIBentryALTinterwordspacing
M.~Antezza, L.~P. Pitaevskii, and S.~Stringari, ``Effect of the casimir-polder
  force on the collective oscillations of a trapped bose-einstein condensate,''
  \emph{Phys. Rev. A}, vol.~70, p. 053619, Nov 2004. [Online]. Available:
  \url{https://link.aps.org/doi/10.1103/PhysRevA.70.053619}
\BIBentrySTDinterwordspacing

\bibitem{Motazedifard2019force}
\BIBentryALTinterwordspacing
A.~Motazedifard, A.~Dalafi, F.~Bemani, and M.~H. Naderi, ``Force sensing in
  hybrid bose-einstein-condensate optomechanics based on parametric
  amplification,'' \emph{Phys. Rev. A}, vol. 100, p. 023815, Aug 2019.
  [Online]. Available:
  \url{https://link.aps.org/doi/10.1103/PhysRevA.100.023815}
\BIBentrySTDinterwordspacing

\bibitem{ratzel_dynamical_2018}
\BIBentryALTinterwordspacing
D.~Rätzel, R.~Howl, J.~Lindkvist, and I.~Fuentes,
  ``\BIBforeignlanguage{en}{Dynamical response of {Bose}–{Einstein}
  condensates to oscillating gravitational fields},''
  \emph{\BIBforeignlanguage{en}{New Journal of Physics}}, vol.~20, no.~7, p.
  073044, Jul. 2018. [Online]. Available:
  \url{https://doi.org/10.1088%2F1367-2630%2Faad272}
\BIBentrySTDinterwordspacing

\bibitem{bravo2020phononic}
\BIBentryALTinterwordspacing
T.~Bravo, D.~Rätzel, and I.~Fuentes, ``Phononic gravity gradiometry with
  bose-einstein condensates, arxiv:2001.10104,'' 2020. [Online]. Available:
  \url{https://arxiv.org/abs/2001.10104}
\BIBentrySTDinterwordspacing

\bibitem{Howl:2019expl}
\BIBentryALTinterwordspacing
R.~Howl, R.~Penrose, and I.~Fuentes, ``Exploring the unification of quantum
  theory and general relativity with a bose{\textendash}einstein condensate,''
  \emph{New Journal of Physics}, vol.~21, no.~4, p. 043047, apr 2019. [Online].
  Available: \url{https://doi.org/10.1088%2F1367-2630%2Fab104a}
\BIBentrySTDinterwordspacing

\bibitem{Sabin:2014gravwave}
\BIBentryALTinterwordspacing
C.~Sab{\'{\i}}n, D.~E. Bruschi, M.~Ahmadi, and I.~Fuentes, ``Phonon creation by
  gravitational waves,'' \emph{New Journal of Physics}, vol.~16, no.~8, p.
  085003, aug 2014. [Online]. Available:
  \url{https://doi.org/10.1088%2F1367-2630%2F16%2F8%2F085003}
\BIBentrySTDinterwordspacing

\bibitem{sabin_thermal_2016}
\BIBentryALTinterwordspacing
C.~Sabín, J.~Kohlrus, D.~E. Bruschi, and I.~Fuentes,
  ``\BIBforeignlanguage{en}{Thermal noise in {BEC}-phononic gravitational wave
  detectors},'' \emph{\BIBforeignlanguage{en}{EPJ Quantum Technology}}, vol.~3,
  no.~1, p.~8, May 2016. [Online]. Available:
  \url{https://doi.org/10.1140/epjqt/s40507-016-0046-4}
\BIBentrySTDinterwordspacing

\bibitem{Schuetzhold:2018int}
\BIBentryALTinterwordspacing
R.~Sch\"utzhold, ``Interaction of a bose-einstein condensate with a
  gravitational wave,'' \emph{Phys. Rev. D}, vol.~98, p. 105019, Nov 2018.
  [Online]. Available:
  \url{https://link.aps.org/doi/10.1103/PhysRevD.98.105019}
\BIBentrySTDinterwordspacing

\bibitem{Robbins_2019}
\BIBentryALTinterwordspacing
M.~P. Robbins, N.~Afshordi, and R.~B. Mann, ``Bose-einstein condensates as
  gravitational wave detectors,'' \emph{Journal of Cosmology and Astroparticle
  Physics}, vol. 2019, no.~07, pp. 032--032, jul 2019. [Online]. Available:
  \url{https://doi.org/10.1088%2F1475-7516%2F2019%2F07%2F032}
\BIBentrySTDinterwordspacing

\bibitem{robbins2021detection}
\BIBentryALTinterwordspacing
M.~P.~G. Robbins, N.~Afshordi, A.~O. Jamison, and R.~B. Mann, ``Detection of
  gravitational waves using parametric resonance in bose-einstein condensates,
  arxiv:2101.03691,'' 2021. [Online]. Available:
  \url{https://arxiv.org/abs/2101.03691}
\BIBentrySTDinterwordspacing

\bibitem{Jin:1996coll}
\BIBentryALTinterwordspacing
D.~S. Jin, J.~R. Ensher, M.~R. Matthews, C.~E. Wieman, and E.~A. Cornell,
  ``Collective excitations of a bose-einstein condensate in a dilute gas,''
  \emph{Phys. Rev. Lett.}, vol.~77, pp. 420--423, Jul 1996. [Online].
  Available: \url{https://link.aps.org/doi/10.1103/PhysRevLett.77.420}
\BIBentrySTDinterwordspacing

\bibitem{Mewes:1996coll}
\BIBentryALTinterwordspacing
M.-O. Mewes, M.~R. Andrews, N.~J. van Druten, D.~M. Kurn, D.~S. Durfee, C.~G.
  Townsend, and W.~Ketterle, ``Collective excitations of a bose-einstein
  condensate in a magnetic trap,'' \emph{Phys. Rev. Lett.}, vol.~77, pp.
  988--991, Aug 1996. [Online]. Available:
  \url{https://link.aps.org/doi/10.1103/PhysRevLett.77.988}
\BIBentrySTDinterwordspacing

\bibitem{Stamper-Kurn:1998coll}
\BIBentryALTinterwordspacing
D.~M. Stamper-Kurn, H.-J. Miesner, S.~Inouye, M.~R. Andrews, and W.~Ketterle,
  ``Collisionless and hydrodynamic excitations of a bose-einstein condensate,''
  \emph{Phys. Rev. Lett.}, vol.~81, pp. 500--503, Jul 1998. [Online].
  Available: \url{https://link.aps.org/doi/10.1103/PhysRevLett.81.500}
\BIBentrySTDinterwordspacing

\bibitem{Katz:2004high}
\BIBentryALTinterwordspacing
N.~Katz, R.~Ozeri, J.~Steinhauer, N.~Davidson, C.~Tozzo, and F.~Dalfovo, ``High
  sensitivity phonon spectroscopy of bose-einstein condensates using
  matter-wave interference,'' \emph{Phys. Rev. Lett.}, vol.~93, p. 220403, Nov
  2004. [Online]. Available:
  \url{https://link.aps.org/doi/10.1103/PhysRevLett.93.220403}
\BIBentrySTDinterwordspacing

\bibitem{Jaskula:2012acoustic}
\BIBentryALTinterwordspacing
J.-C. Jaskula, G.~B. Partridge, M.~Bonneau, R.~Lopes, J.~Ruaudel, D.~Boiron,
  and C.~I. Westbrook, ``Acoustic analog to the dynamical casimir effect in a
  bose-einstein condensate,'' \emph{Phys. Rev. Lett.}, vol. 109, p. 220401, Nov
  2012. [Online]. Available:
  \url{https://link.aps.org/doi/10.1103/PhysRevLett.109.220401}
\BIBentrySTDinterwordspacing

\bibitem{stamper1999excitation}
D.~Stamper-Kurn, A.~Chikkatur, A.~G{\"o}rlitz, S.~Inouye, S.~Gupta,
  D.~Pritchard, and W.~Ketterle, ``Excitation of phonons in a bose-einstein
  condensate by light scattering,'' \emph{Physical review letters}, vol.~83,
  no.~15, p. 2876, 1999.

\bibitem{Schley:2013planck}
\BIBentryALTinterwordspacing
R.~Schley, A.~Berkovitz, S.~Rinott, I.~Shammass, A.~Blumkin, and J.~Steinhauer,
  ``Planck distribution of phonons in a bose-einstein condensate,'' \emph{Phys.
  Rev. Lett.}, vol. 111, p. 055301, Jul 2013. [Online]. Available:
  \url{https://link.aps.org/doi/10.1103/PhysRevLett.111.055301}
\BIBentrySTDinterwordspacing

\bibitem{brennecke2007cavity}
\BIBentryALTinterwordspacing
F.~Brennecke, T.~Donner, S.~Ritter, T.~Bourdel, M.~K{\"o}hl, and T.~Esslinger,
  ``Cavity qed with a bose--einstein condensate,'' \emph{nature}, vol. 450, no.
  7167, pp. 268--271, 2007. [Online]. Available:
  \url{https://doi.org/10.1038/nature06120}
\BIBentrySTDinterwordspacing

\bibitem{brennecke2008cavity}
\BIBentryALTinterwordspacing
F.~Brennecke, S.~Ritter, T.~Donner, and T.~Esslinger, ``Cavity optomechanics
  with a bose-einstein condensate,'' \emph{Science}, vol. 322, no. 5899, pp.
  235--238, 2008. [Online]. Available:
  \url{https://doi.org/10.1126/science.1163218}
\BIBentrySTDinterwordspacing

\bibitem{Note1}
This approximation is valid if the parameters and the geometry of the setup are
  chosen such that the coupling of the modes in the elongated direction to
  those in the transverse directions is sufficiently suppressed, for example,
  in the case of very tight transverse confinement (as considered below in the
  example application).

\bibitem{Pitaevskii:2003bose}
L.~P. Pitaevskii and S.~Stringari, \emph{{Bose--Einstein} condensation}.\hskip
  1em plus 0.5em minus 0.4em\relax Clarendon Press - Oxford, 2003.

\bibitem{gaunt2013bose}
\BIBentryALTinterwordspacing
A.~L. Gaunt, T.~F. Schmidutz, I.~Gotlibovych, R.~P. Smith, and Z.~Hadzibabic,
  ``Bose-einstein condensation of atoms in a uniform potential,'' \emph{Phys.
  Rev. Lett.}, vol. 110, p. 200406, May 2013. [Online]. Available:
  \url{https://link.aps.org/doi/10.1103/PhysRevLett.110.200406}
\BIBentrySTDinterwordspacing

\bibitem{Note2}
Note that $a_\protect \mathrm {sc}$ can, in general, be widely tuned for some
  atom species that possess Feshbach resonances (e.g. $^{23}$Na \cite
  {inouye1998observation}, $^{85}$Rb \cite {roberts1998resonant} and $^{87}$Rb
  \cite {marte2002feshbach}) by employing strong magnetic fields to modify the
  s-wave scattering length. Unfortunately, also the three-body loss rate is
  strongly enhanced near a Feshbach resonance (see also \cite
  {chin2010feshbach}), where three-body loss is the dominating loss mechanism
  for trapped condensates and the dominating limitating factor for the maximal
  experimental time considered in this paper. Therefore, Feshbach resonances
  are of little use for our proposal and will not be considered.

\bibitem{Salasnich2002eff}
\BIBentryALTinterwordspacing
L.~Salasnich, A.~Parola, and L.~Reatto, ``Effective wave equations for the
  dynamics of cigar-shaped and disk-shaped bose condensates,'' \emph{Phys. Rev.
  A}, vol.~65, p. 043614, Apr 2002. [Online]. Available:
  \url{https://link.aps.org/doi/10.1103/PhysRevA.65.043614}
\BIBentrySTDinterwordspacing

\bibitem{Note3}
For $\rho _\protect \mathrm {1d} a_\protect \mathrm {sc} \gtrsim 1$, we would
  need to replace $\protect \hat {\psi }^\dagger \protect \hat {\psi }$ by more
  complicated functions of $\protect \hat {\psi }^\dagger \protect \hat {\psi
  }$ (see e.g. \cite {Salasnich2002eff,rauer2018recurrences}). For the sake of
  simplicity we refrain from this here.

\bibitem{nagy2009nonlinear}
\BIBentryALTinterwordspacing
D.~Nagy, P.~Domokos, A.~Vukics, and H.~Ritsch, ``Nonlinear quantum dynamics of
  two bec modes dispersively coupled by an optical cavity,'' \emph{The European
  Physical Journal D}, vol.~55, no.~3, pp. 659--668, 2009. [Online]. Available:
  \url{https://doi.org/10.1140/epjd/e2009-00265-7}
\BIBentrySTDinterwordspacing

\bibitem{ritsch2013cold}
\BIBentryALTinterwordspacing
H.~Ritsch, P.~Domokos, F.~Brennecke, and T.~Esslinger, ``Cold atoms in
  cavity-generated dynamical optical potentials,'' \emph{Rev. Mod. Phys.},
  vol.~85, pp. 553--601, Apr 2013. [Online]. Available:
  \url{https://link.aps.org/doi/10.1103/RevModPhys.85.553}
\BIBentrySTDinterwordspacing

\bibitem{aspelmeyer2014cavity}
\BIBentryALTinterwordspacing
M.~Aspelmeyer, T.~J. Kippenberg, and F.~Marquardt, ``Cavity optomechanics,''
  \emph{Reviews of Modern Physics}, vol.~86, no.~4, p. 1391, 2014. [Online].
  Available: \url{http://dx.doi.org/10.1103/RevModPhys.86.1391}
\BIBentrySTDinterwordspacing

\bibitem{pikovski:thesis}
\BIBentryALTinterwordspacing
I.~Pikovski, ``Macroscopic quantum systems and gravitational phenomena,'' Ph.D.
  dissertation, uniwien, 2014. [Online]. Available:
  \url{http://othes.univie.ac.at/33500/}
\BIBentrySTDinterwordspacing

\bibitem{Note4}
This is the effect of the refractive index change due to the presence of the
  condensate atoms in the cavity.

\bibitem{demkowicz2015quantum}
\BIBentryALTinterwordspacing
R.~Demkowicz-Dobrza{\'n}ski, M.~Jarzyna, and J.~Ko{\l}ody{\'n}ski, ``Quantum
  limits in optical interferometry,'' \emph{Progress in Optics}, vol.~60, pp.
  345--435, 2015. [Online]. Available:
  \url{https://doi.org/10.1016/bs.po.2015.02.003}
\BIBentrySTDinterwordspacing

\bibitem{Tozzo:2004pho}
\BIBentryALTinterwordspacing
C.~Tozzo and F.~Dalfovo, ``Phonon evaporation in freely expanding
  {Bose--Einstein} condensates,'' \emph{Phys. Rev. A}, vol.~69, p. 053606, May
  2004. [Online]. Available:
  \url{https://link.aps.org/doi/10.1103/PhysRevA.69.053606}
\BIBentrySTDinterwordspacing

\bibitem{Andrews:1996dir}
\BIBentryALTinterwordspacing
M.~R. Andrews, M.-O. Mewes, N.~J. van Druten, D.~S. Durfee, D.~M. Kurn, and
  W.~Ketterle, ``Direct, nondestructive observation of a {Bose} condensate,''
  \emph{Science}, vol. 273, no. 5271, pp. 84--87, 1996. [Online]. Available:
  \url{http://science.sciencemag.org/content/273/5271/84}
\BIBentrySTDinterwordspacing

\bibitem{Stamper:1998col}
\BIBentryALTinterwordspacing
D.~M. Stamper-Kurn, H.-J. Miesner, S.~Inouye, M.~R. Andrews, and W.~Ketterle,
  ``Collisionless and hydrodynamic excitations of a {Bose--Einstein}
  condensate,'' \emph{Phys. Rev. Lett.}, vol.~81, pp. 500--503, Jul 1998.
  [Online]. Available:
  \url{https://link.aps.org/doi/10.1103/PhysRevLett.81.500}
\BIBentrySTDinterwordspacing

\bibitem{Schuetzhold2006det}
\BIBentryALTinterwordspacing
R.~Sch\"utzhold, ``Detection scheme for acoustic quantum radiation in
  bose-einstein condensates,'' \emph{Phys. Rev. Lett.}, vol.~97, p. 190405, Nov
  2006. [Online]. Available:
  \url{https://link.aps.org/doi/10.1103/PhysRevLett.97.190405}
\BIBentrySTDinterwordspacing

\bibitem{schumm2005matter}
\BIBentryALTinterwordspacing
T.~Schumm, S.~Hofferberth, L.~M. Andersson, S.~Wildermuth, S.~Groth,
  I.~Bar-Joseph, J.~Schmiedmayer, and P.~Kr{\"u}ger, ``Matter-wave
  interferometry in a double well on an atom chip,'' \emph{Nature physics},
  vol.~1, no.~1, pp. 57--62, 2005. [Online]. Available:
  \url{https://doi.org/10.1038/nphys125}
\BIBentrySTDinterwordspacing

\bibitem{schweigler2017experimental}
\BIBentryALTinterwordspacing
T.~Schweigler, V.~Kasper, S.~Erne, I.~Mazets, B.~Rauer, F.~Cataldini,
  T.~Langen, T.~Gasenzer, J.~Berges, and J.~Schmiedmayer, ``Experimental
  characterization of a quantum many-body system via higher-order
  correlations,'' \emph{Nature}, vol. 545, no. 7654, pp. 323--326, 2017.
  [Online]. Available: \url{https://doi.org/10.1038/nature22310}
\BIBentrySTDinterwordspacing

\bibitem{van2018projective}
\BIBentryALTinterwordspacing
Y.~D. Van~Nieuwkerk, J.~Schmiedmayer, and F.~Essler, ``Projective phase
  measurements in one-dimensional bose gases,'' \emph{SciPost Phys.}, vol.~5,
  p.~46, 2018. [Online]. Available:
  \url{https://scipost.org/10.21468/SciPostPhys.5.5.046}
\BIBentrySTDinterwordspacing

\bibitem{gluza2020quantum}
\BIBentryALTinterwordspacing
M.~Gluza, T.~Schweigler, B.~Rauer, C.~Krumnow, J.~Schmiedmayer, and J.~Eisert,
  ``Quantum read-out for cold atomic quantum simulators,'' \emph{Communications
  Physics}, vol.~3, no.~1, pp. 1--8, 2020. [Online]. Available:
  \url{https://doi.org/10.1038/s42005-019-0273-y}
\BIBentrySTDinterwordspacing

\bibitem{vanner2011pulsed}
\BIBentryALTinterwordspacing
M.~R. Vanner, I.~Pikovski, G.~D. Cole, M.~Kim, {\v{C}}.~Brukner, K.~Hammerer,
  G.~J. Milburn, and M.~Aspelmeyer, ``Pulsed quantum optomechanics,''
  \emph{Proceedings of the National Academy of Sciences}, vol. 108, no.~39, pp.
  16\,182--16\,187, 2011. [Online]. Available:
  \url{https://www.pnas.org/content/108/39/16182}
\BIBentrySTDinterwordspacing

\bibitem{Giorgini:1998dam}
S.~{Giorgini}, ``{Damping in dilute {Bose} gases: A mean-field approach},''
  \emph{\pra}, vol.~57, pp. 2949--2957, Apr. 1998.

\bibitem{Szepfalusy:1974on}
\BIBentryALTinterwordspacing
P.~Szépfalusy and I.~Kondor, ``On the dynamics of continuous phase
  transitions,'' \emph{Annals of Physics}, vol.~82, no.~1, pp. 1 -- 53, 1974.
  [Online]. Available:
  \url{http://www.sciencedirect.com/science/article/pii/0003491674903303}
\BIBentrySTDinterwordspacing

\bibitem{Shi:1998fin}
\BIBentryALTinterwordspacing
H.~Shi and A.~Griffin, ``Finite-temperature excitations in a dilute
  bose-condensed gas,'' \emph{Physics Reports}, vol. 304, no. 1–2, pp. 1 --
  87, 1998. [Online]. Available:
  \url{http://www.sciencedirect.com/science/article/pii/S0370157398000155}
\BIBentrySTDinterwordspacing

\bibitem{Fedichev:1998damp}
\BIBentryALTinterwordspacing
P.~O. Fedichev, G.~V. Shlyapnikov, and J.~T.~M. Walraven, ``Damping of
  low-energy excitations of a trapped {Bose--Einstein} condensate at finite
  temperatures,'' \emph{Phys. Rev. Lett.}, vol.~80, pp. 2269--2272, Mar 1998.
  [Online]. Available:
  \url{http://link.aps.org/doi/10.1103/PhysRevLett.80.2269}
\BIBentrySTDinterwordspacing

\bibitem{yuen2015enhanced}
\BIBentryALTinterwordspacing
B.~Yuen, I.~Barr, J.~Cotter, E.~Butler, and E.~Hinds, ``Enhanced oscillation
  lifetime of a bose--einstein condensate in the 3d/1d crossover,'' \emph{New
  Journal of Physics}, vol.~17, no.~9, p. 093041, 2015. [Online]. Available:
  \url{https://doi.org/10.1088/1367-2630/17/9/093041}
\BIBentrySTDinterwordspacing

\bibitem{mazets2011dynamics}
\BIBentryALTinterwordspacing
I.~E. Mazets, ``Dynamics and kinetics of quasiparticle decay in a
  nearly-one-dimensional degenerate bose gas,'' \emph{Phys. Rev. A}, vol.~83,
  p. 043625, Apr 2011. [Online]. Available:
  \url{https://link.aps.org/doi/10.1103/PhysRevA.83.043625}
\BIBentrySTDinterwordspacing

\bibitem{Burt:1997coh}
\BIBentryALTinterwordspacing
E.~A. Burt, R.~W. Ghrist, C.~J. Myatt, M.~J. Holland, E.~A. Cornell, and C.~E.
  Wieman, ``Coherence, correlations, and collisions: What one learns about
  bose-einstein condensates from their decay,'' \emph{Phys. Rev. Lett.},
  vol.~79, pp. 337--340, Jul 1997. [Online]. Available:
  \url{https://link.aps.org/doi/10.1103/PhysRevLett.79.337}
\BIBentrySTDinterwordspacing

\bibitem{Takuso:2003spin}
\BIBentryALTinterwordspacing
Y.~Takasu, K.~Maki, K.~Komori, T.~Takano, K.~Honda, M.~Kumakura, T.~Yabuzaki,
  and Y.~Takahashi, ``Spin-singlet {Bose--Einstein} condensation of
  two-electron atoms,'' \emph{Phys. Rev. Lett.}, vol.~91, p. 040404, Jul 2003.
  [Online]. Available:
  \url{https://link.aps.org/doi/10.1103/PhysRevLett.91.040404}
\BIBentrySTDinterwordspacing

\bibitem{mehta_three-body_2007}
\BIBentryALTinterwordspacing
N.~P. Mehta, B.~D. Esry, and C.~H. Greene, ``Three-body recombination in one
  dimension,'' \emph{Physical Review A}, vol.~76, no.~2, p. 022711, Aug. 2007.
  [Online]. Available:
  \url{https://link.aps.org/doi/10.1103/PhysRevA.76.022711}
\BIBentrySTDinterwordspacing

\bibitem{haller_three-body_2011}
\BIBentryALTinterwordspacing
E.~Haller, M.~Rabie, M.~J. Mark, J.~G. Danzl, R.~Hart, K.~Lauber, G.~Pupillo,
  and H.-C. Nägerl, ``Three-{Body} {Correlation} {Functions} and
  {Recombination} {Rates} for {Bosons} in {Three} {Dimensions} and {One}
  {Dimension},'' \emph{Physical Review Letters}, vol. 107, no.~23, p. 230404,
  Dec. 2011. [Online]. Available:
  \url{https://link.aps.org/doi/10.1103/PhysRevLett.107.230404}
\BIBentrySTDinterwordspacing

\bibitem{tolra_observation_2004}
\BIBentryALTinterwordspacing
B.~L. Tolra, K.~M. O’Hara, J.~H. Huckans, W.~D. Phillips, S.~L. Rolston, and
  J.~V. Porto, ``Observation of {Reduced} {Three}-{Body} {Recombination} in a
  {Correlated} 1d {Degenerate} {Bose} {Gas},'' \emph{Physical Review Letters},
  vol.~92, no.~19, p. 190401, May 2004. [Online]. Available:
  \url{https://link.aps.org/doi/10.1103/PhysRevLett.92.190401}
\BIBentrySTDinterwordspacing

\bibitem{raetzel2021decay}
\BIBentryALTinterwordspacing
D.~R\"atzel and R.~Sch\"utzhold, ``Decay of quantum sensitivity due to
  three-body loss in bose-einstein condensates,'' \emph{Phys. Rev. A}, vol.
  103, p. 063321, Jun 2021. [Online]. Available:
  \url{https://link.aps.org/doi/10.1103/PhysRevA.103.063321}
\BIBentrySTDinterwordspacing

\bibitem{Howl_2017}
\BIBentryALTinterwordspacing
R.~Howl, C.~Sab{\'{\i}}n, L.~Hackermüller, and I.~Fuentes, ``Quantum
  decoherence of phonons in bose{\textendash}einstein condensates,''
  \emph{Journal of Physics B: Atomic, Molecular and Optical Physics}, vol.~51,
  no.~1, p. 015303, nov 2017. [Online]. Available:
  \url{https://doi.org/10.1088%2F1361-6455%2Faa9622}
\BIBentrySTDinterwordspacing

\bibitem{Note5}
Note that the conventional mode decomposition for uniform condensates would be
  modes of the form $\protect \qopname \relax o{exp}(i \hbar k_n z)$ and their
  complex conjugate. However, since the light mode is confined in a cavity, the
  light-atom interaction is symmetric under the exchange of the propagation
  direction of the quasiparticles and modes of the form $\protect \qopname
  \relax o{cos}(k_{n}(z+L/2))$ and $\protect \qopname \relax
  o{sin}(k_{n}(z+L/2))$ are directly addressed. The explicit coupling strength
  will depend on the position of the condensate inside the optical resonator.

\bibitem{Egorov:2013meas}
\BIBentryALTinterwordspacing
M.~Egorov, B.~Opanchuk, P.~Drummond, B.~V. Hall, P.~Hannaford, and A.~I.
  Sidorov, ``Measurement of $s$-wave scattering lengths in a two-component
  {Bose--Einstein} condensate,'' \emph{Phys. Rev. A}, vol.~87, p. 053614, May
  2013. [Online]. Available:
  \url{http://link.aps.org/doi/10.1103/PhysRevA.87.053614}
\BIBentrySTDinterwordspacing

\bibitem{Note6}
Compare Eq. (\ref {eq:Pn}) with Eq. (\ref {eq:Udisp}).

\bibitem{Note7}
Note that we are interested in keeping the photon number sufficiently large to
  reduce the potential effects of photon number fluctuations (shot noise). The
  photon number can be increased, for example, by increasing the detuning. For
  sufficiently high photon numbers, the photon number fluctuations will average
  out due to the short lifetime of photons in the cavity of much less than the
  read-out time of 400\protect \,ns.

\bibitem{Note8}
In this case, momentum transfer to the phonons would couple photons out of the
  cavity mode which implies a different coupling Hamiltonian than the one used
  in this article.

\bibitem{Lorenzo_2014}
\BIBentryALTinterwordspacing
L.~A.~D. Lorenzo and K.~C. Schwab, ``Superfluid optomechanics: coupling of a
  superfluid to a superconducting condensate,'' \emph{New Journal of Physics},
  vol.~16, no.~11, p. 113020, nov 2014. [Online]. Available:
  \url{https://doi.org/10.1088/1367-2630/16/11/113020}
\BIBentrySTDinterwordspacing

\bibitem{inouye1998observation}
\BIBentryALTinterwordspacing
S.~Inouye, M.~Andrews, J.~Stenger, H.-J. Miesner, D.~M. Stamper-Kurn, and
  W.~Ketterle, ``Observation of feshbach resonances in a bose--einstein
  condensate,'' \emph{Nature}, vol. 392, no. 6672, pp. 151--154, 1998.
  [Online]. Available: \url{https://doi.org/10.1038/32354}
\BIBentrySTDinterwordspacing

\bibitem{roberts1998resonant}
\BIBentryALTinterwordspacing
J.~Roberts, N.~Claussen, J.~P. Burke~Jr, C.~H. Greene, E.~A. Cornell, and
  C.~Wieman, ``Resonant magnetic field control of elastic scattering in cold
  rb85,'' \emph{Phys. Rev. Lett.}, vol.~81, pp. 5109--5112, Dec 1998. [Online].
  Available: \url{https://link.aps.org/doi/10.1103/PhysRevLett.81.5109}
\BIBentrySTDinterwordspacing

\bibitem{marte2002feshbach}
\BIBentryALTinterwordspacing
A.~Marte, T.~Volz, J.~Schuster, S.~D{\"u}rr, G.~Rempe, E.~Van~Kempen, and
  B.~Verhaar, ``Feshbach resonances in rubidium 87: Precision measurement and
  analysis,'' \emph{Phys. Rev. Lett.}, vol.~89, p. 283202, Dec 2002. [Online].
  Available: \url{https://link.aps.org/doi/10.1103/PhysRevLett.89.283202}
\BIBentrySTDinterwordspacing

\bibitem{chin2010feshbach}
\BIBentryALTinterwordspacing
C.~Chin, R.~Grimm, P.~Julienne, and E.~Tiesinga, ``Feshbach resonances in
  ultracold gases,'' \emph{Rev. Mod. Phys.}, vol.~82, pp. 1225--1286, Apr 2010.
  [Online]. Available:
  \url{https://link.aps.org/doi/10.1103/RevModPhys.82.1225}
\BIBentrySTDinterwordspacing

\bibitem{safranek2015}
\BIBentryALTinterwordspacing
D.~\v{S}afr\'{a}nek, , A.~R. Lee, and I.~Fuentes, ``Quantum parameter
  estimation using multi-mode gaussian states,'' \emph{New Journal of Physics},
  vol.~17, p. 073016, 2015. [Online]. Available:
  \url{https://doi.org/10.1088/1367-2630/17/7/073016}
\BIBentrySTDinterwordspacing

\bibitem{koklovett_qprocessing}
P.~Kok and B.~W. Lovett, \emph{Introduction to optical quantum information
  processing}.\hskip 1em plus 0.5em minus 0.4em\relax Cambridge University
  Press, 2010.

\bibitem{Note9}
Note that, for $\protect \bar \psi _0=\psi _0e^{-i\mu t}$, we would have
  $i\hbar \partial _t\protect \bar \psi _0=\mu $ and recover the stationary GP
  equation (\ref {eq:statgp}).

\bibitem{minguzzi2004numerical}
\BIBentryALTinterwordspacing
A.~Minguzzi, S.~Succi, F.~Toschi, M.~Tosi, and P.~Vignolo, ``Numerical methods
  for atomic quantum gases with applications to bose--einstein condensates and
  to ultracold fermions,'' \emph{Physics reports}, vol. 395, no. 4-5, pp.
  223--355, 2004. [Online]. Available:
  \url{https://doi.org/10.1016/j.physrep.2004.02.001}
\BIBentrySTDinterwordspacing

\bibitem{barenghi2016primer}
C.~F. Barenghi and N.~G. Parker, \emph{A primer on quantum fluids}.\hskip 1em
  plus 0.5em minus 0.4em\relax Springer, 2016.

\end{thebibliography}

\appendix

\section{Cavity frequency shift}
\label{sec:freqshift}

Due to the proportionality of $\hat{H}_\rm{disp}$ in Eq. (\ref{eq:Hdisp}) to the photon number operator $\hat{a}^\dagger\hat{a}$, we can conclude that $\hat{H}_\rm{disp}$ leads to a frequency shift of the cavity frequency proportional to the number of atoms $N_0$ in the atomic ensemble weighted with the overlap of $\hat{\psi}^\dagger(z)\hat{\psi}(z)$ and the cavity mode square. Effectively, this is a refractive index change due to the presence of the atoms in the cavity. We take the average frequency shift into account from the start by renormalizing the cavity mode frequency as $\omega_c \rightarrow \omega_c - \delta \omega_c$, where
\begin{equation}
	\delta \omega_c = \frac{g_0^2 N_0}{\Delta_A} \int dz \,\chi_\rm{BT}(z) f_\rm{cav}(z)^2/L\,,
\end{equation} 
where $L$ its length of the box potential and $\chi_\rm{BT}$ is the characteristic function of the 1-dimensional box potential in the $z$-direction (i.e. $\chi_\rm{BT} = 1$ inside and $\chi_\rm{BT}=0$ outside the box, respectively). For each single run of the experiment, the atom number is fixed and does not have quantum properties. Therefore, we can formally replace $N_0$ with the atom number operator $\hat N = \mathcal{A}\int dz\, \hat\psi^\dagger(z) \hat\psi(z)$ in the following.

Then, together with the Hamiltonian governing the dynamics of the atomic ensemble, the total Hamiltonian of our system is
\begin{align}\label{eq:hamiltonianshifted}
    \nonumber \hat{H}_\text{total} & = \hbar \omega_c \hat{a}^\dagger\hat{a} + \int dz \, \hat{\psi}^\dagger(z)  \, \hbar\frac{g_0^2}{\Delta_A} \,\hat{a}^\dagger\hat{a} \left( f_\rm{cav}^2(z)  -  \int dz'\, \chi_\rm{BT}(z') f_\rm{cav}^2(z')/L\right) \hat{\psi}(z)   \\
    &  +  \int dz\,\hat{\psi}^\dagger(z)\Bigg[ -\frac{\hbar^2}{2m}\partial_z^2 +  \frac{\tilde{g}}{2} \hat{\psi}^\dagger(z)\hat{\psi}(z)  + V_0\left(z\right) +  \delta V_\rm{ext}\left(z,t\right) \Bigg]\hat{\psi}(z) 
\end{align}

\section{The interaction Hamiltonian}
\label{sec:intham}

Starting from the split of the atomic field into the ground state part and excitations as $\hat\psi = \hat{c}_0 \psi_0 + \hat{\varphi}$, neglecting all contributions of the perturbation $\hat\varphi$ and the fluctuations of the light field $\delta\hat{a}$, we obtain
\begin{eqnarray}
	\nonumber  \hat H_\rm{total} &=& \int dz\, \hat{c}_0^\dagger \psi_0^{*}\left(-\frac{\hbar^2}{2m} \partial_z^2  + V + \frac{\tilde{g}}{2}\hat{c}_0^\dagger\hat{c}_0|\psi_0|^2\right)\hat{c}_0\psi_0  \,.
\end{eqnarray}
With the canonical commutation relations $[\hat{c}_0,\hat{c}_0^\dagger]=1$, the Heisenberg equation of motion for the ground state leads to
\begin{equation}
	\hat{c}_0(t) = \hat{c}_0(0) \exp\left(-\frac{i}{\hbar}\left(\mu t + \int_0^t dt'\delta \mu_\rm{osc}(t') + \tilde{g}t \int dz\,|\psi_0|^4 \left(\hat{N}_0 - (N_0+1)\right)\right)\right)\,,
\end{equation}
where $\hat{N}_0=\hat{c}_0^\dagger\hat{c}_0$
for normalized $\psi_0$, where we defined $\delta \mu_\rm{osc}(t)= \int dz\, |\psi_0|^2 V_\rm{osc}(t)$. If we assume that the state of condensate is restricted to the particle sector around the particle number $N_0\gg 1$ that is much larger than the particle number fluctuations, we can neglect the last term in the above equation and obtain
\begin{equation}
	\hat{c}_0(t) \approx \hat{c}_0(0) \exp\left(-\frac{i}{\hbar}\left(\mu t + \int_0^t dt'\delta \mu_\rm{osc}(t')\right)\right)\,,
\end{equation}
With this observation, we re-define the splitting of the atomic field operator in the Heisenberg picture as
\begin{eqnarray}
	\hat\psi= (\hat c_0 \psi_0 + \hat\vartheta) e^{-i\left( \mu t + \int_0^t dt'\, \delta\mu_\rm{osc}(t')\right)/\hbar} \,.
\end{eqnarray} 
Then, we find that the time evolution of $\hat\psi' :=\hat c_0 \psi_0 + \hat\vartheta$ is governed by the Heisenberg equation with respect to the Hamiltonian
\begin{equation}
	\hat H_\text{total}':= \hat H_\text{total} - (\mu+\delta\mu_\rm{osc}) \hat N
\end{equation} 
(similar to the grand canonical Hamiltonian) where $\hat N = \int dz\, \hat\psi^{\prime\dagger}\hat\psi^\prime$ is the full atom number operator. $\mu+\delta\mu_\rm{osc}$ can be interpreted as a time-dependent chemical potential.

We continue with the Hamiltonian $\hat{H}'_\rm{total}$ and use the Bogoliubov approximation $\hat{c}_0\rightarrow \sqrt{N_0}\mathbb{I}$ to find the expansion up to second order in $\hat\vartheta$
\begin{eqnarray}\label{eq:expansionfull}
	\nonumber  \hat H'_\rm{total} &=& -\hbar\Delta_c \delta\hat{a}^\dagger\delta\hat{a} + N_0\int dz\, \psi_0^{*}\left(-\frac{\hbar^2}{2m} \partial_z^2  + V_0 + \overline{\delta V} + \frac{\tilde{g}N_0}{2}|\psi_0|^2\right)\psi_0  \\
	\nonumber && + \sqrt{N_0}\int dz\, \left(\hat\vartheta^{\dagger}\left(-\frac{\hbar^2}{2m} \partial_z^2  + V_0 + \overline{\delta V} + \tilde{g}N_0|\psi_0|^2 \right)\psi_0 + h.c.\right) \\
	\nonumber && + \int dz\, \left(\hat\vartheta^{\dagger}\left(-\frac{\hbar^2}{2m} \partial_z^2  + V_0 + \overline{\delta V} + 2\tilde{g}N_0|\psi_0|^2 \right)\hat\vartheta + \frac{\tilde{g}N_0}{2}\left(\hat\vartheta^{\dagger 2}\psi_0^{ 2} + \psi_0^{*2}\hat\vartheta^{2}\right)\right)
	\\
	&& + \sqrt{N_0} \int dz\, \tilde{g}\left(\hat\vartheta^{\dagger 2}\hat\vartheta \psi_0 + \psi_0^{*}\hat\vartheta^{\dagger}\hat\vartheta^{2}\right) + \int dz\, \frac{\tilde{g}}{2}\hat\vartheta^{\dagger 2}\hat\vartheta^{2}\\
	\nonumber && +    \hbar\frac{g_0^2\sqrt{N_{ph}} }{\Delta_A} \left(\delta\hat{a}^\dagger  + \delta\hat{a}  \right)  \int dz \, f_\rm{cav}^2\left( N_0 \psi_0^{*} \psi_0 +  \sqrt{N_0}\left(\psi_0^{*}\hat\vartheta +  \psi_0\hat\vartheta^\dagger \right) +  \hat\vartheta^\dagger\hat\vartheta \right)\\
	\nonumber && + N_0 \int dz\, \psi_0^{*} V_\rm{osc}\psi_0  + \sqrt{N_0}\int dz\, \left(\psi_0^{*} V_\rm{osc}\hat\vartheta + \hat\vartheta^{\dagger} V_\rm{osc}\psi_0\right) +  \int dz\, \hat\vartheta^{\dagger} V_\rm{osc}\hat\vartheta \\
	\nonumber && - N_0 \mu \int dz\, \psi_0^{*} \psi_0 - N_0 \delta\mu_\rm{osc} \int dz\, \psi_0^{*} \psi_0 - \sqrt{N_0} \mu \int dz\, \left(\psi_0^{*} \hat\vartheta + \hat\vartheta^{\dagger}\psi_0\right)  \\
	\nonumber &&  - \sqrt{N_0}\delta\mu_\rm{osc} \int dz\, \left(\psi_0^{*} \hat\vartheta + \hat\vartheta^{\dagger}\psi_0\right) - \mu\int dz\, \hat\vartheta^{\dagger} \hat\vartheta- \delta\mu_\rm{osc}\int dz\, \hat\vartheta^{\dagger} \hat\vartheta
\end{eqnarray}
In the last three lines, we see the contribution of the time dependent potential perturbation and $ - (\mu+\delta\mu_\rm{osc}) \hat N$. The second term in the second to last line and the first term in the third to last line cancel. 

With Eq. (\ref{eq:statgp}), the second term in the first line in Eq. (\ref{eq:expansionfull}) gives the classical energy of the condensate, and with the first term in the second last line of Eq. (\ref{eq:expansionfull})
\begin{equation}
	 E^{(0)} = -\frac{\tilde{g}N_0^2}{2}\int dz\, |\psi_0|^4\,.
\end{equation} 
Again with the stationary GP equation (\ref{eq:statgp}), the second line of Eq. (\ref{eq:expansionfull}) becomes
\begin{equation}
	\hat H^{(1)} =  \sqrt{N_0}\mu\int dz\, \left(\hat\vartheta^{\dagger}\psi_0 + h.c.\right)\,,
\end{equation} 
which cancels with the last term in the second last line of Eq. (\ref{eq:expansionfull}). The third line of Eq. (\ref{eq:expansionfull}) gives rise to the Bogoliubov Hamiltonian. We combine the third line and the second term in the last line of Eq. (\ref{eq:expansionfull}) as
\begin{eqnarray}
	\hat H^{(2)} &:=& \quad \int dz\, \left(\hat\vartheta^{\dagger}\left(-\frac{\hbar^2}{2m} \partial_z^2  + V_0 + \overline{\delta V} - \mu + 2\tilde{g}N_0|\psi_0|^2 \right)\hat\vartheta \right. \\
	\nonumber && \left. + \frac{\tilde{g}N_0}{2}\left(\hat\vartheta^{\dagger 2}\psi_0^{2} + \psi_0^{*2}\hat\vartheta^{2}\right)\right) \,,
\end{eqnarray} 

As explained in the main text, we expand the field operator in terms of Bogoliubov modes describing the quasiparticle excitations as
\begin{equation}
	\hat\vartheta = \sum_n  \left(u_n\hat b_{n} + v_n^*\hat b_{n}^\dagger \right)\,,
\end{equation}
where $[b_{n},b_{m}^\dagger]=\delta_{nm}$, the mode functions $u_n$ and $v_n$ fulfill the stationary Bogoliubov-de~Gennes (BDG) equations 
\begin{eqnarray}\label{eq:BDG}
	\hbar \omega_n u_n(z) &=& \left(-\frac{\hbar^2}{2m} \partial_z^2  + V_0 + \overline{\delta V} - \mu + 2\tilde{g}N_0|\psi_0|^2\right)u_n(z) + \tilde{g}N_0\psi_0^2 v_n(z)\\
	-\hbar \omega_n v_n(z) &=& \left(-\frac{\hbar^2}{2m} \partial_z^2  + V_0 + \overline{\delta V} - \mu + 2\tilde{g}N_0|\psi_0|^2\right)v_n(z) + \tilde{g}N_0\psi_0^{*2} u_n(z)\,
\end{eqnarray}
and are normalized with respect to the inner product
\begin{equation}\label{eq:normauv}
	\int dz \, \left(u_n^* u_m - v_n^* v_m\right) = \delta_{nm}\,.
\end{equation}
The expansion of $\hat\vartheta$ in the Bogoliubov basis diagonalizes the Bogoliubov Hamiltonian as $\hat H_\rm{BdG} = \, :H^{(2)}:\, = \sum_n \hbar \omega_n \hat b_{n}^\dagger \hat b_{n}$, where $:\quad:$ denotes the normal ordering with respect to the Bogoliubov mode operators $b_{n}^\dagger$ and $\hat b_{n}$, which leads to the omission of the constant vacuum energy. 

The fifth line in equation (\ref{eq:expansionfull}) governs the back action on the light field. The first term in brackets corresponds to a time-independent shift of the cavity field frequency. This is discussed in more details in the main text and in Appendix \ref{sec:freqshift} and we omit this term. The third term in brackets is of higher order and will be omitted as well. The second term gives rise to the back action Hamiltonian in equation (\ref{eq:backaction_hamiltonian_unexp}). Furthermore, we combine the remaining terms of Eq. (\ref{eq:expansionfull}) to the driving Hamiltonian in equation (\ref{eq:Hprimeint}) and neglect terms of third and fourth order in $\hat\vartheta$ that give rise to quasiparticle-quasiparticle interactions that are discussed in Section \ref{sec:damping}.

\section{Derivation of the driving Hamiltonian due to an external force gradient}
\label{sec:derivationintHamPi}

Starting from Eqs. (\ref{eq:interaction_hamiltonian}) and (\ref{eq:coefficients}), to first order in $1/\sqrt{N_0}$ and with $V_\rm{osc}=-z^2 G_0 \sin\left(\Omega t\right)$, we obtain the driving Hamiltonian due to the external force gradient as
\begin{align}  
        \nonumber \hat{H}_\rm{int} &:= \hat{H}_\rm{dr} =\sum_{n} \left(P_{n} \hat{b}_{n} e^{-i\omega_{n}t}  + P_{n}^* \hat{b}_{n}^{\dagger}e^{i\omega_{n}t}\right)
\end{align}
where 
\begin{align}
      P_{n} = \sqrt{N_0}   \int dz \,  (V_\rm{osc} - \delta\mu_\rm{osc}) \,\left( \psi_0^* u_n + \psi_0 v_n \right)  \,.
\end{align}
Making the assumptions of Sec. \ref{sec:applications} (homogeneous condensate in a box trap etc.), assuming the resonance condition $\Omega=\omega_{n_\rm{low}}$ and applying the RWA, we find the interaction Hamiltonian
\begin{align}  
        \nonumber \hat{H}_{int} &=  \Pi_{n_\rm{low}} \left(\hat{b}_{n_\rm{low}} - \hat{b}_{n_\rm{low}}^{\dagger}\right)\,,
\end{align}
where
\begin{align}
       \Pi_{n_\rm{low}} = i\frac{G_0 \sqrt{N_0}}{2\sqrt{L}}   \int dz \,  \left(z^2 - \frac{L^2}{12}\right) \,\left( u_{n_\rm{low}} + v_{n_\rm{low}} \right)  = i\frac{G_0 \sqrt{N_0}}{2L \sigma_{n_\rm{low}}}   \int dz \,  \left(z^2 - \frac{L^2}{12}\right) \,\varphi_{n_\rm{low}}^c\,.
\end{align}

\section{Ground state and mode function perturbations}
\label{sec:gstate_pert}

In this appendix, we discuss the dependence of the ground state on the time-averaged perturbation of the external potential $\delta V$ that is discussed in Sec. \ref{sec:hamiltonian}. We have that
$\delta \psi_0$ is a solution of the linearized GP equation
\begin{equation}\label{eq:statgp_lin}
	\overline{\delta V}\,\bar\psi_0 + \left(-\frac{\hbar^2}{2m} \partial_z^2  + 2\tilde{g}N_0|\bar{\psi}_0|^2 \right)\delta\psi_0 + \tilde{g}N_0\bar{\psi}_0^2 \delta\psi_0^* = \mu \delta\psi_0\,.
\end{equation}
We apply the Thomas-Fermi approximation and neglect the kinetic term. Then, the linearized GP equation can be solved as
\begin{equation}
	\rm{Re}\left(\delta\psi_0\right)  = - \overline{\delta V}/(2\mu \mathcal{V}^{1/2})\,,
\end{equation}
and $\rm{Im}(\delta\psi_0) = 0$. In a similar fashion, we can find the modifications of the quasiparticle mode function $u_n$ and $v_n$. As all of these modifications are of first order in $\overline{\delta V}$ and $\hat{H}_\rm{dr}$ as well as $\hat{H}_\rm{ba}$ are both of first order in the external potential and light field interaction, respectively, their modifications due to $\overline{\delta V}$ will be of higher order and can be neglected for the purposes of this article.

\section{Fundamental sensitivity of the pulsed readout}
\label{app:fund-readout-limit}

When considering only the resonant targeted quasiparticle mode for the pulsed measurement scheme described in section \ref{sec:pulsed}, the time-evolution operator has the form
\begin{equation}
    \hat{U}_{pulse}=\exp\left[-\frac{i}{\hbar} \kappa_{2n_\rm{cav}} \left(\sqrt{N_{ph}} + \left(\delta\hat{a}+\delta\hat{a}^{\dagger}\right)\right)\left(\hat{b}_{2n_\rm{cav}}+\hat{b}_{2n_\rm{cav}}^{\dagger}\right)\Delta t \right].
\end{equation}
The measurement sequence comprising two mirror operations and the gravitationally induced displacement can be described with the time-evolution operator
\begin{equation}\hat{U}_{D}=\left(\hat{U}_{n,c}^{M}\right)^{\dagger}\hat{D}_{n}\hat{U}_{n,c}^{M}
\end{equation}
where $\hat{D}_{n}=\exp\left[-i\mathcal{P}_{n}\left(\hat{b}_{n}-\hat{b}_{n}^{\dagger}\right)\right]$, $\mathcal{P}_{n}=\Pi_{n}t/\hbar$, and $\hat{U}_{n,c}^{M}$ is given by (\ref{eq:mirror}) for the specific modes labelled $n$ and $c$ with the interaction time appropriately tuned. For some initial density matrix $\rho_{0}$, the final state of the condensate-cavity system is then described as
\begin{equation}
    \rho_{F}=\hat{U}_{pulse}\hat{U}_{D}\rho_{0}\hat{U}_{D}^{\dagger}\hat{U}_{pulse}^{\dagger}.
\end{equation}
For notational convenience, we also define $\chi=\kappa_{2n_\rm{cav}}\Delta t/\hbar$. We define the operator basis $\hat{X}_{cav}=\delta\hat{a}+\delta\hat{a}^{\dagger}$ and $\hat{P}_{cav}=i\left(\delta\hat{a}^{\dagger}-\delta\hat{a}\right)$ and the vector $\hat{x}=\left(\hat{X}_{cav},\hat{P}_{cav}\right)$. For an initially thermal quasiparticle state (negligible initial occupancy in the high order mode c) and a coherent cavity state, the displacement vector in this basis for the reduced cavity state is given by
\begin{equation}
    d=\left\langle \hat{x}\right\rangle =\left(0,4\chi\mathcal{P}_{n}\right)
\end{equation}
where the expectation values are taken with respect to $\rho_F$. The covariance matrix, with elements defined by $\Sigma_{i,j}=\left\langle \hat{x}_{i}\hat{x}_{j}+\hat{x}_{j}\hat{x}_{i}\right\rangle -\left\langle \hat{x}_{i}\right\rangle \left\langle \hat{x}_{j}\right\rangle$, is given by
\begin{equation}
    \Sigma=\begin{pmatrix}1 & 0\\
    0 & 1+4\chi^{2}
    \end{pmatrix}.
\end{equation}
The quantum Fisher information (QFI) $H_{\rho}\left(\lambda\right)$ gives a measure of the amount of information about a parameter $\lambda$ which can be extracted from a state $\rho$ optimised over all possible measurements. The QFI can be calculated for Gaussian states using only the displacement vector and covariance matrix \cite{safranek2015}. We find that the QFI for estimating the displacement $\mathcal{P}_{n}$ of the quasiparticle state is given by
\begin{equation}
    H_{\rho_{F}}\left(\mathcal{P}_{n}\right)=\frac{16\chi^{2}}{1+4\chi^{2}}.
\end{equation}
The quantum Cramer-Rao bound \cite{koklovett_qprocessing} links the possible measurement sensitivity with the QFI, and in our case it has the form
\begin{equation}
    \left(\Delta\mathcal{P}_{n}\right)^{2}=\frac{1}{N_{meas}}\left(\frac{1}{4}+\frac{1}{16\chi^{2}}\right)
\end{equation}
where $N_{meas}$ is the number of repetitions of the measurement scheme performed.

\section{A toy model for high-energy modes - trapezoid potential}
\label{app:toymodel}

To show how the coefficients $A^c_n$ and $A^s_n$ of modes in the high-energy regime are constructed and vary with the mode number, we consider a toy model. We start by modelling the trap potential as a well with finitely steep walls, i.e. with a certain inclination, such that
\begin{equation}
    V_0(z) = a( -(z+L/2) \Theta(-(z+L/2)) 
    + (z-L/2) \Theta(z-L/2))\,.
\end{equation}
We want the potential to rise to the kinetic energy of the high energy move in a fraction of $L$, i.e. $a=b\hbar^2(2k_\rm{cav})^2/(2mL)$, where $b\gtrsim 10$. For high-energy modes, we can set $v_n=0$, and for $\delta V=0$, the BDG equations in (\ref{eq:BDG}) become the stationary Schrödinger equation for $u_n$ 
\begin{eqnarray}\label{eq:stSchröd}
 0 &=& \left(-\frac{\hbar^2}{2m} \partial_z^2  + (\tilde{V}_0(z)-	E_n) \right)u_n(z)
\end{eqnarray}
with $E_n=\hbar\omega_n$ and the potential $\tilde{V}_0(z)= V_0 - \mu_0 + 2\tilde{g}N_0|\psi_0|^2$. We note that for a small healing length in comparison to the trap length, we can approximate $2\tilde{g}N_0|\psi_0|^2\approx 2\tilde{g}\rho_\rm{1d} = 2\mu_0$ and the interaction energy $\mu_0 = \tilde{g}\rho_\rm{1d}$ can be regarded as a small offset of the potential such that
\begin{eqnarray}
    \nonumber \tilde{V}_0(z) &:=& V_0(z) - \tilde{g}\rho_\rm{1d} \\
    \nonumber &\approx& a( -(z+L/2-\mu_0/a) \Theta(-(z+L/2))    + (z-L/2+\mu_0/a) \Theta(z-L/2)) \\
    && + (\Theta(z+L/2) - \Theta(z-L/2)) \mu_0\,.
\end{eqnarray}
and $L+2\mu_0/a \approx L$. Then, we define three regions I, II and III, where I and III are left of $z=-L/2$ and right of $z=L/2$, respectively, and II lies in between I and III. 
\begin{figure}[b]
\centering
\includegraphics[width=8cm,angle=0]{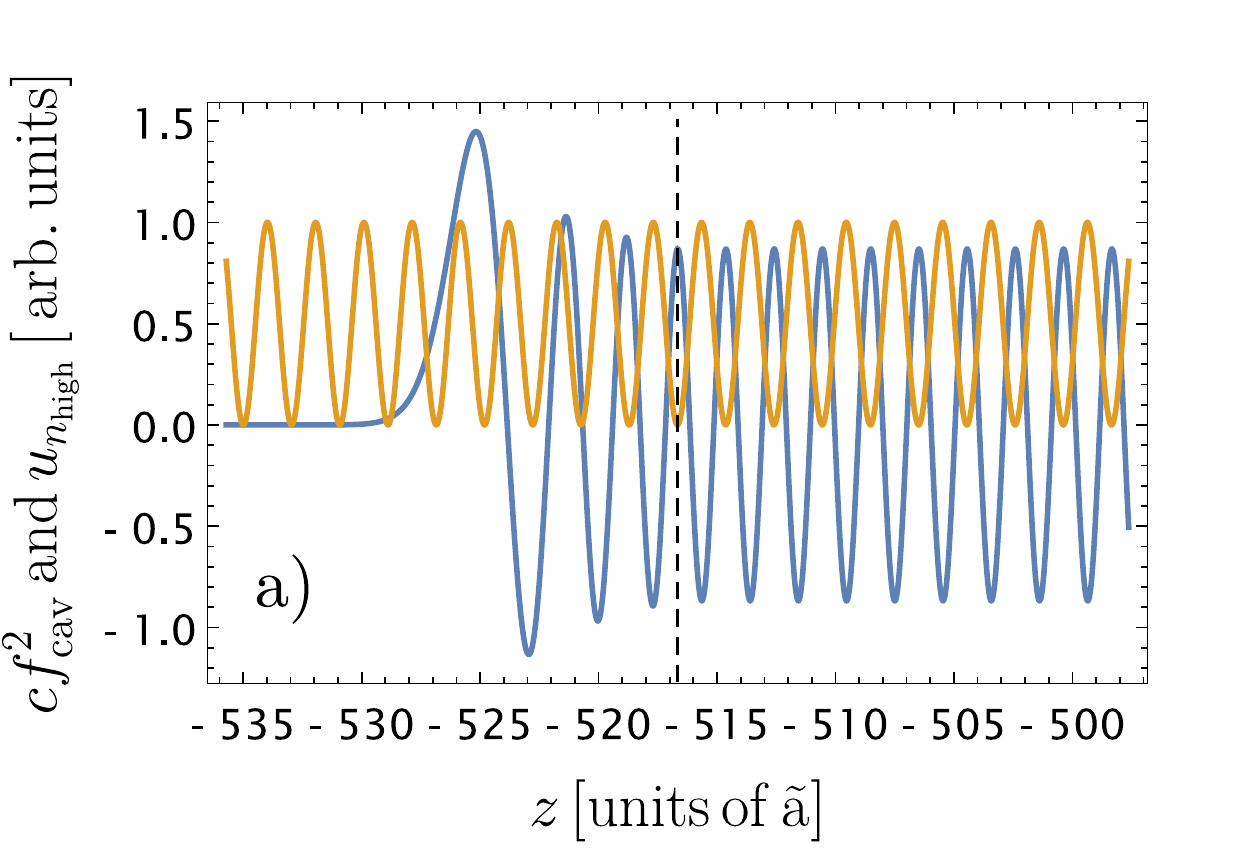}
\includegraphics[width=8cm,angle=0]{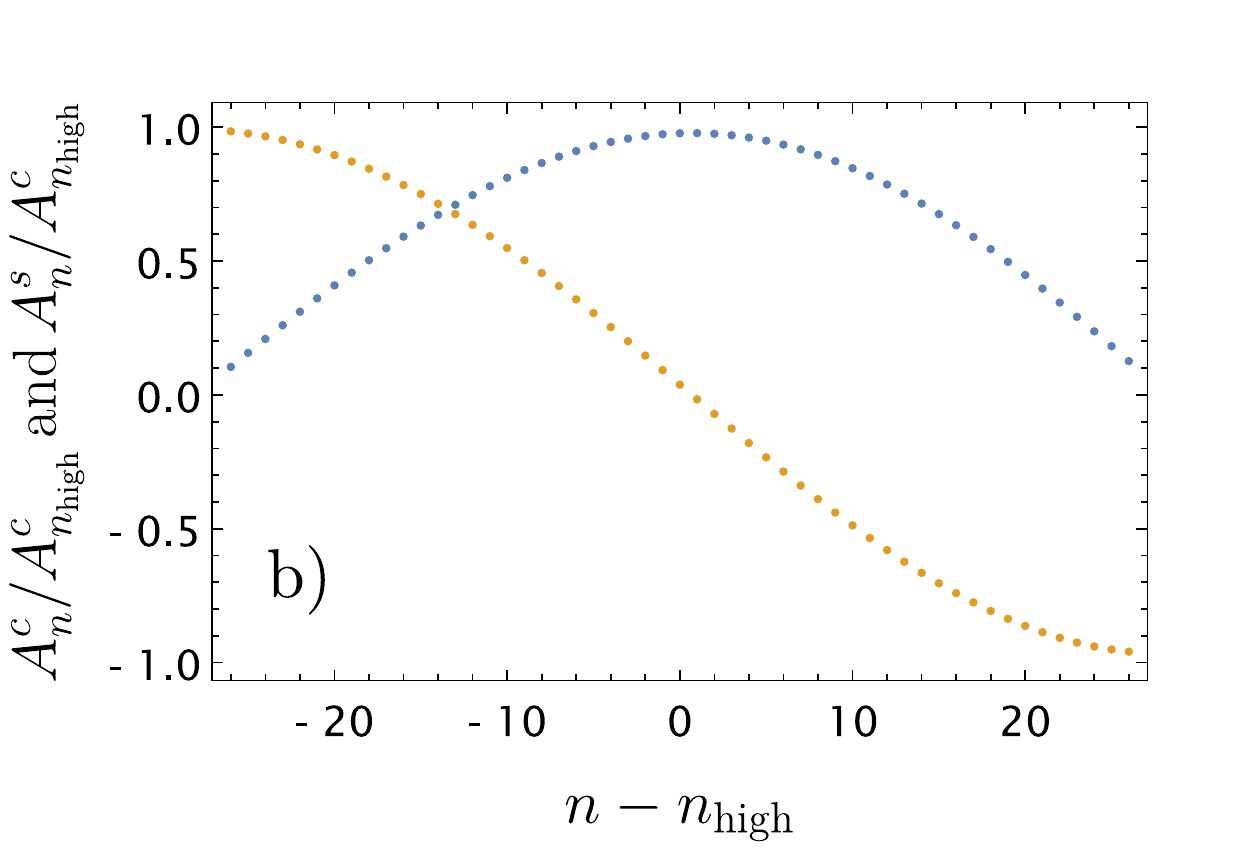}
\includegraphics[width=8cm,angle=0]{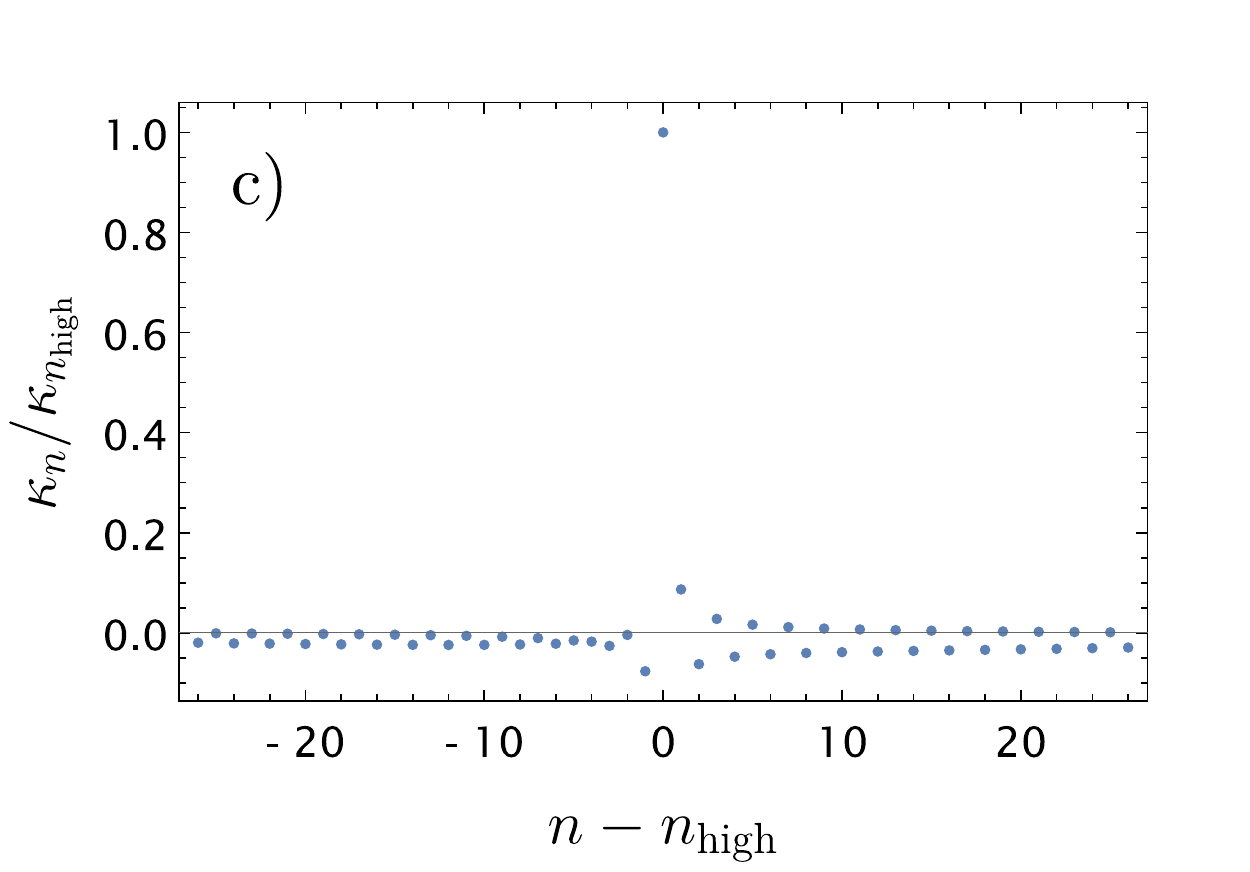}
\caption{\label{fig:trapezoid-potential} {\small Plot of various functions for similar parameters as those considered in Section \ref{sec:applications} and $b=108.5$:  a) A plot of the spatial dependence of the square of the optical mode $f_\rm{cav}^2$ multiplied by an arbitrary factor $c$ (orange line) and the quasiparticle mode $u_{n_\rm{high}}$ with $n_\rm{high}=1027$ (blue line) close to $z=-L/2$ (marked by the vertical black dashed line). The overlap between these functions is very large up to a small region left of $z=-L/2$ where the quasiparticle mode reaches a maximum before it quickly decays. b) A plot of the coefficients $A_n^c$ (blue dots) and $A_n^s$ (orange dots) of the mode in region II. For values $n$ close to $n_\rm{high}$, the $A_n^s$ coefficient almost vanishes while $A_n^c$ is close to 1. c) Coupling coefficient $\kappa_n$ for modes close to $n=n_\rm{high}$.   }}
\end{figure}

In region III, we find the stationary Schrödinger equation
\begin{eqnarray}
 0 &=& \left(- \partial_z^2  + a\frac{2m}{\hbar^2}(z - L/2 + \mu_0/a - E_n/a) \right)u_n(z)
\end{eqnarray}
We redefine the spatial coordinate as $\bar{z}=(2m a/\hbar^2)^{1/3}(z - L/2 + \mu_0/a - E_n/a)$ such that
\begin{eqnarray}\label{eq:stSchrödI}
 0 &=& \left(-\partial_{\bar{z}}^2  + \bar{z} \right)\bar{u}_n(\bar{z})\,.
\end{eqnarray}
This differential equation is solved by the Airy function $\rm{Ai}(\bar{z})$ and we obtain the solution
\begin{equation}
    \tilde{u}_{n,I}(\tilde{z}) = C_{R,n} \rm{Ai}(\tilde{z} - \tilde{L}/2 + \tilde{E}_n)\,.
\end{equation} 
where we defined $\tilde{z}=\tilde{a}^{1/3} z$, $\tilde{L}=\tilde{a}^{1/3} L$, $\tilde{E}_n=\tilde{a}^{-2/3} 2m(\mu_0 - E_n)/\hbar^2$ and $\tilde{a}=2m a/\hbar^2=b(2k_\rm{cav})^2/L$. We find the corresponding solution in region I by reflection at the origin and for the region II, we find the differential equation
\begin{eqnarray}\label{eq:stSchrödII}
 0 &=& \left(-\partial_{\tilde{z}}^2  + \tilde{E}_n \right)\tilde{u}_n(\tilde{z})
\end{eqnarray}
which is simply solved by
\begin{equation}
    \tilde{u}_{n,II}(\tilde{z}) = \sqrt{\frac{2}{L}} \left(A^\rm{c}_n \cos\left(\tilde{k}_{n}\left(\tilde{z}+\frac{\tilde{L}}{2}\right)\right) + A^\rm{s}_n \sin\left(\tilde{k}_{n}\left(\tilde{z}+\frac{\tilde{L}}{2}\right)\right)\right)
\end{equation}
where $\tilde{k}_{n}=\sqrt{\tilde{E}_n}=\tilde{a}^{-1/3} k_n$. From imposing the continuity of the wave function and its first derivative at $z=-L/2$ and $z=L/2$, we obtain the following expressions for the coefficients
\begin{eqnarray}
    \nonumber C_{R,n} &=& \mathcal{N} \left( \cos(\tilde{k}_{n} \tilde{L}) - \frac{\rm{Ai}'(-\tilde{E}_n)}{\tilde{k}_{n} Ai(-\tilde{E}_n)}\sin(\tilde{k}_{n} \tilde{L}) \right)\\
             A^\rm{c}_n &=& \sqrt{\frac{L}{2}} \mathcal{N} \rm{Ai}(-\tilde{E}_n) \\
    \nonumber A^\rm{s}_n &=& -\sqrt{\frac{L}{2}} \mathcal{N} \frac{\rm{Ai}'(-\tilde{E}_n)}{\tilde{k}_n}\,.
\end{eqnarray}
and the consistency condition
\begin{equation}\label{eq:spectrumcond}
    \rm{Ai}'(-\tilde{E}_n)\cos(\tilde{k}_n\tilde{L}) + \frac{\tilde{E}_n \rm{Ai}(-\tilde{E}_n)^2 -\rm{Ai}'(-\tilde{E}_n)^2}{\tilde{k}_n \rm{Ai}(-\tilde{E}_n)} \sin(\tilde{k}_n\tilde{L}) = 0
\end{equation}
which defines the spectrum of possible $\tilde{E}_n$. The factor $\mathcal{N}$ is a normalization constant. Solutions to equation (\ref{eq:spectrumcond}) as well as $\mathcal{N}$ can be found numerically. 

In figure \ref{fig:trapezoid-potential}, we show the squared magnitude of the cavity mode function and the quasiparticle mode function with $k_{n_\rm{high}}\sim 2k_\rm{cav}$, a plot of the coefficients $A_n^c$ and $A_n^s$ as well as the coupling coefficient $\kappa_n$ for $n$ close to $n_\rm{high}$. We plot these for parameters equivalent to those used in Section \ref{sec:applications}, besides the trap length which is chosen to be 198 microns and $b=108.5$ which does not appear in the main text. We find that the length scale $\tilde{a}^{-1/3}\sim 190\,$nm. We have chosen the values for $L$ and $b$ such that $A_n^c/A_n^s \gg 1$ and $k_{n_\rm{high}}$ is very close to $2k_\rm{cav}$ to optimize the overlap. In particular, we see that the coupling to modes with $n\neq n_\rm{high}$ is very small and can be neglected which was assumed in the main text.

\section{Numerical treatment of the GP equation}
\label{app:numericalGP}

In this appendix, we introduce the numerical method used for the calculation of the ground state presented in Fig. \ref{fig:ground-state}. 
We restrict our considerations to one-dimensional quasi-condensates and start with the time dependent GP equation
\begin{equation}\label{eq:time_dep_gp}
	\left(-\frac{\hbar^2}{2m} \partial_z^2  + V_0 + g_\rm{1d} N_0 |\bar\psi|^2 \right)\bar\psi = i\hbar \partial_t\bar\psi\,
\end{equation}
for the ground state wave function normalized as $\int dz\,|\bar\psi|^2 = 1$ \footnote{Note that, for $\bar\psi_0=\psi_0e^{-i\mu t}$, we would have $i\hbar\partial_t\bar\psi_0=\mu$ and recover the stationary GP equation (\ref{eq:statgp}).}. We define $\tilde{z}=z/\xi$, $\tau = \mu_0 t/\hbar$, $\tilde{V}=V/\mu_0$, $\tilde{\psi}=\sqrt{\xi}\bar\psi$ and $\mu_0 =g_\rm{1d}\rho_\rm{0,1d}$, where the one-dimensional density $\rho_\rm{0,1d}=N_0/L$ and the healing length $\xi=\hbar/\sqrt{4m_a \mu_0}$ have been already defined in the main text. Then, equation (\ref{eq:time_dep_gp}) can be rewritten in dimensionless form as
\begin{equation}\label{eq:time_dep_gp_dimless}
	\left(-2\partial_{\tilde{z}}^2  + \tilde{V} + \tilde{L}|\tilde\psi|^2 \right)\tilde\psi = i \partial_\tau \tilde\psi\,,
\end{equation}
where $\tilde{L}=L/\xi$ and $\int d\tilde{z}\,|\tilde\psi|^2 = 1$. 

To obtain the ground state, we perform an imaginary time propagation \cite{minguzzi2004numerical}, that is, we use imaginary time steps $-i\,d\tau$. Furthermore, we solve the differential equation numerically on a spatial grid with a discrete time split step method (see \cite{minguzzi2004numerical} and Appendix A of \cite{barenghi2016primer}). More concretely, the wave function is represented as a one-dimensional array $\tilde\psi_\rm{ar}=\{\tilde\psi_j\}$ of complex values of $\tilde\psi$ at the points of a one-dimensional array $\{\tilde{z}_j\}$ with $J$ entries of equidistant points from an interval of length $L_g$ of the $\tilde{z}$-axis. In each time step from $\tau_s$ to $\tau_{s+1}=\tau_s+d\tau$, first, we perform the operation $U^{1/2}_{\tilde{z}}(\tau_s,d\tau)=\{\exp(-i(\tilde{V}(\tilde{z}_j,\tau_s) + \tilde{L}|\tilde\psi_j(\tau_s)|^2)d\tau/2)\}$, on the wave function $\tilde\psi_\rm{ar}(\tau_s)$, perform a discrete Fourier transform $\mathcal{F}$ of $U^{1/2}_{\tilde{z}}(\tau_s,d\tau)\tilde\psi_\rm{ar}(\tau_s)$ where Fourier space is represented by another array $\{\tilde{k}_j\}$ of wave numbers $k_j=2\pi (j-1-J/2)/L_g$. Then, we apply the operation $U_{\tilde{k}}(d\tau)=\exp(-2i \tilde{k}^2 d\tau)$ on $\mathcal{F}[U^{1/2}_{\tilde{z}}(\tau_s,d\tau)\tilde\psi_\rm{ar}(\tau_s)]$, apply the Fourier back-transform $\mathcal{F}^{-1}$ and another time $U^{1/2}_{\tilde{z}}(\tau_s,d\tau)$. That is, the full operation can be written as
 \begin{equation}
    \tilde{\psi}_\rm{ar}(\tau_{s+1}) = U^{1/2}_{\tilde{z},0}(\tau_s,-i\,d\tau) \mathcal{F}^{-1}[U_{\tilde{k}}(-i\,d\tau)\mathcal{F}[U^{1/2}_{\tilde{z},0}(\tau_s,-i\,d\tau)\tilde\psi_\rm{ar}(\tau_s)]]\,,
\end{equation}
where $U^{1/2}_{\tilde{z},0}$ is equivalent to $U^{1/2}_{\tilde{z}}$ with $\tilde{V}$ replaced by the time-independent trap potential $\tilde{V}_0=V_0/\mu_0$ and $U_{\tilde{k}}=U^{1/2}_{\tilde{k}}U^{1/2}_{\tilde{k}}$. After each time step, $\tilde{\psi}_\rm{ar}$ is normalized to one.

\end{document}